\documentclass[fleqn,usenatbib]{mnras}
\usepackage{newtxtext,newtxmath}
\usepackage[T1]{fontenc}
\DeclareRobustCommand{\VAN}[3]{#2}
\let\VANthebibliography\thebibliography
\def\thebibliography{\DeclareRobustCommand{\VAN}[3]{##3}\VANthebibliography}

\usepackage{siunitx}
\usepackage{mathtools}
\usepackage{physics}

\usepackage{graphicx}
\usepackage[FIGTOPCAP]{subfigure}
\usepackage{rotating}
\usepackage{epstopdf}
\usepackage{dcolumn}
\usepackage{bm}
\usepackage{float}
\usepackage{xcolor}

\newcommand\msun{\, \rm M_\odot}

\newcommand{\eq}[1]{\begin{align}#1\end{align}}
\newcommand{\subeq}[1]{\begin{subequations}\begin{align}#1\end{align}\end{subequations}}

\definecolor{purple}{rgb}{0.59, 0.44, 0.84}

\newcommand{\de}{\mathrm{d}}

\def\msun{\mathrm{M}_\odot}

\def\bsub{\begin{subequations}}
\def\esub{\end{subequations}}

\def\mb{m_\mathrm{b}}

\def\rate{\mathcal{R}}

\def\barr{\begin{eqnarray}}
\def\earr{\end{eqnarray}}
\def\bsub{\begin{subequations}}
\def\esub{\end{subequations}}

\title[Binaries in stellar clusters]{Orbital evolution of LIGO/Virgo binaries in stellar clusters driven by cluster tides, stellar encounters and general relativity}

\author[A. Rasskazov \& R. R. Rafikov]{
Alexander Rasskazov$^1$ \& Roman R. Rafikov$^{1,2}\thanks{E-mail: rrr@damtp.cam.ac.uk}$
\\
$^1$Department of Applied Mathematics and Theoretical Physics, University of Cambridge, Wilberforce Road, Cambridge CB3 0WA, UK\\
$^2$Institute for Advanced Study, Einstein Drive, Princeton, NJ 08540, USA
}

\date{Accepted XXX. Received YYY; in original form ZZZ}

\pubyear{2023}

\begin{document}
\label{firstpage}
\pagerange{\pageref{firstpage}--\pageref{lastpage}}
\maketitle

\begin{abstract}
Origin of LIGO/Virgo gravitational wave events may involve production of binaries with relativistic components in dense stellar systems --- globular or nuclear star clusters --- and their subsequent evolution towards merger. Orbital parameters of these binaries (the inner orbit) and their motion inside the cluster (the outer orbit) evolve due to both external agents --- random encounters with cluster stars and cluster tides due to the smooth cluster potential --- and the internal ones --- various sources of dissipation and precession within the binary. We present a numerical framework --- Binary Evolution in Stellar Clusters (BESC) --- that follows the evolution of the binary inner and outer orbits accounting for all these effects simultaneously, enabling efficient Monte Carlo studies. The secular effect of cluster tides is computed in the singly-averaged approximation, without averaging over the outer binary orbit. As to stellar encounters, we include the effects of both close and distant flybys on the inner and outer orbits of the binary, respectively. In particular, this allows us to explicitly account for the dynamical friction sinking the binary towards the cluster centre. Also, given our focus on the LIGO/Virgo sources, we include the general relativistic precession (which suppresses cluster tides at high eccentricities) and the gravitational wave emission (shrinking the binary orbit). We use BESC to illustrate a number of characteristic binary evolutionary outcomes and discuss relative contributions of different physical processes. BESC can also be used to study other objects in clusters, e.g. blue stragglers, hot Jupiters, X-ray binaries, etc. 
\end{abstract}

\begin{keywords}
gravitation -- gravitational wave -- celestial mechanics -- binaries: general -- stars: kinematics and dynamics -- galaxies: star clusters: general
\end{keywords}


\section{Introduction}
\label{sec:intro}


At the moment of writing, 76 compact binary mergers have been detected by the LIGO/VIRGO/KAGRA collaboration \citep{LIGO}, with 70 of them classified as black hole-black\footnote{In this work we will be using ``BH binary'' and ``compact object binary'' interchangeably.} hole (BH-BH) mergers.  For these mergers to occur in the first place, some dynamical mechanism must exist that could bring the binary components close enough for the merger timescale due to the gravitational wave (GW) emission to be at least shorter than the Hubble time. In particular, for a circular $10\msun+10\msun$ binary the semimajor axis would have to be $a\lesssim\SI{0.09}{AU}$. 

The specific mechanism responsible for reducing the compact binary semi-major axis depends on the environment in which the binary resides. In the galactic field, the compact binaries can be naturally produced as a result of stellar evolution in isolated massive binaries via the common envelope evolution \citep{Belczynski2016}; the compact binary then exists in isolation and can only shrink its orbit via the GW emission. Another scenario in the field involves hierarchical triples that may shrink their orbits via the Lidov-Kozai eccentricity oscillations accompanied by the GW emission at the pericentre \citep{Silsbee2017}. 

In denser environments, i.e. in star clusters, compact binaries can either be primordial or form in close triple or quadruple stellar encounters. Their mergers in the cluster could be facilitated by the occasional close flybys of stars or other compact objects, which may harden the binary, i.e. reduce its semi-major axis, until the GW emission takes over \citep{Rodriguez2016,Fragione2018}. Such close encounters with other stars in a cluster can be interpreted as resulting from the {\it granularity} of the gravitational potential inside the cluster.

Recently, \citet{paper1,paper2,paper3} have proposed another mechanism for reducing the semi-major axis of compact binaries in star clusters. They pointed out that under certain conditions the tidal forces arising due to the large-scale inhomogeneity of the {\it smooth} component of the cluster potential are able to drive binary eccentricity to high values, at which point the GW emission could become effective at reducing the binary semi-major axis. This mechanism is similar to the Lidov-Kozai effect with the tidal field of the whole cluster playing a role of the gravitational perturbation due to the third body. Using this idea, \citet{paper3} have carried out a simplified Monte Carlo calculation of the merger rate of compact binaries due to the cluster tides. However, this work ignored all dynamical effects of encounters with other stars in the cluster that must always be present in some level. 

Given the number and complexity of physical processes affecting the orbital evolution of binaries (not necessarily consisting of compact objects) in clusters, their dynamics has often been explored using numerical models with different levels of approximation. For example, the Cluster Monte Carlo Code \citep{cmc} is a statistical framework that considers only the effects of stellar flybys by randomly changing the energy and angular momentum of every member of a cluster at every timestep. While that method accounts for effects such as (nonresonant) relaxation and close encounters, it ignores the effect of distant encounters (with pericentre distances much higher than the binary semimajor axis) or the cluster tidal field on the inner orbital parameters of the binary. Many other studies also consider only the effects of encounters with cluster stars on the binary orbital elements, for example \citet{HamersTremaine}\footnote{That paper actually considers a hot Jupiter orbiting around a star rather than a BH binary.} and \citet{distantEncountersEffect}. At the other extreme, \citet{paper3} and \citet{BubPetrovich} modeled exclusively the effects of cluster tides on binary orbits. Full N-body simulations of stellar clusters naturally take into account both the smooth and spatially-fluctuating components of the cluster potential, but they are usually computationally expensive \citep{Li2023}.

The goal of this work is to present a novel numerical framework --- Binary Evolution in Stellar Clusters\footnote{BESC and the relevant supporting documentation are available at ~~~~~ \url{https://github.com/RZCas/binary-evolution-in-a-cluster}.} (BESC) --- for efficiently following the evolution of the orbital elements of an individual binary on time intervals as long as the Hubble time, by simultaneously accounting for the effects of the smooth and fluctuating components of the cluster potential, as well as the general relativistic effects. We describe the details of the physical and numerical implementations of this method and illustrate them with several individual examples, deferring to future work the use of BESC for statistical studies of populations of binaries in stellar clusters. Also, one of the goals of BESC is to help us understand the relative importance of cluster tides and stellar encounters for the orbital evolution of a binary and to determine the conditions under which it can become very eccentric.

Our paper is organized as follows. After discussing in \S\ref{sec:physics} the relative roles of different cluster-specific processes in binary evolution, we describe the implementation of the key ingredients of BESC in \S\ref{section:Physics}: encounters (\S\ref{section:encounters}), cluster tides (\S\ref{section:tidalEffects}), and GR effects (\S\ref{section:GREffects}). We then provide a comparison of our singly-averaged implementation of cluster tides in BESC with the existing doubly-averaged results (\S\ref{section:SAvsDA}), and then describe some representative outcomes of binary evolution in clusters (\S\ref{section:examples}). We discuss our results in \S\ref{sec:disc} (including the preliminary statistics of outcomes in \S\ref{sec:stat_outcomes}) and conclude in \S\ref{section:conclusions}. Finally, Appendix \ref{appendix} is devoted to motivating the parameter choices used in BESC.

\begin{table*}
\caption{Comparison of the physical processes accounted for in different models of a binary evolution in a stellar cluster}
\centering
\begin{tabular}{|c| c c c c c c|}
\hline
Paper & Outer orbit & Cluster tides & Cluster evolution & Dynamical friction & Strong encounters & Weak encounters \\ [0.5ex] 
\hline
\citet{cmc} & approx. & no & yes & yes & yes & no \\
\hline
\citet{distantEncountersEffect} & no & no & no & no & approx. & yes \\
\hline
\citet{HamersTremaine} & no & no & no & no & yes & yes\\
\hline
\citet{BubPetrovich} & yes & yes & no & no & no & no \\
\hline
\citet{paper3} & yes & yes & no & no & no & no \\
\hline
This work & yes & yes & no & yes & yes & yes \\
\hline
\end{tabular}
\label{table:comparison}
\end{table*}


\section{Physical processes affecting binary evolution}
\label{sec:physics}


Binaries orbiting in dense stellar systems are subject to a range of physical processes that may affect the dynamical characteristics of the binaries. Some of these processes affect only the inner orbit of the binary, i.e. the motion of the binary components around its centre of mass (CoM), while others affect the outer orbit of the binary, i.e. motion of the CoM within the cluster, or both. 

The outer orbit of the binary within the cluster is fully determined by the cluster potential $\Phi({\bf r})$ and the initial conditions --- position and velocity of the CoM in the cluster. However, because of the random flybys of cluster stars past the binary its outer orbit also evolves, see \S\ref{section:encounters}. The effect of these stellar encounters can be qualitatively decomposed into
\begin{itemize}
\item The dynamical friction due to the smooth density field of the cluster --- a secular effect (\S\ref{section:DF}),
\item Random changes of the binary velocity in its orbit, equivalent to stochastic changes of the initial conditions for the subsequent integration of the outer orbit.
\end{itemize}

At the same time, the inner orbit of the binary evolves due to the following effects:
\begin{itemize}
\item Close encounters with cluster stars (as for the outer orbit), which can change not only the eccentricity and inclination but also the semi-major axis of the binary (\S\ref{section:encounters}),
\item Secular evolution of the inner orbit due to cluster tides --- inhomogeneity of the smooth component of the cluster potential --- leaving the binary semi-major axis unchanged (\S\ref{section:tidalEffects}),
\item General relativistic (GR) effects: orbital precession and GW emission (\S\ref{section:GREffects}).
\end{itemize}

Additional possible evolutionary factor is the variation of the cluster properties on long timescales due to non-resonant and resonant relaxation. However, in this work we will neglect overall cluster evolution for simplicity, to better focus on other aspects of binary dynamics. 

Table \ref{table:comparison} provides a comparison of the physical processes accounted for in our present work and in other existing studies of the binary evolution in stellar clusters. It will be further discussed in \S\ref{section:comparison}. The implementation of the various aforementioned physical effects will be described in \S\ref{section:EvolutionOfTheOuterOrbit}, while a brief discussion of their relative importance for binary evolution is provided next.

\begin{figure*}
	\includegraphics[width=\textwidth]{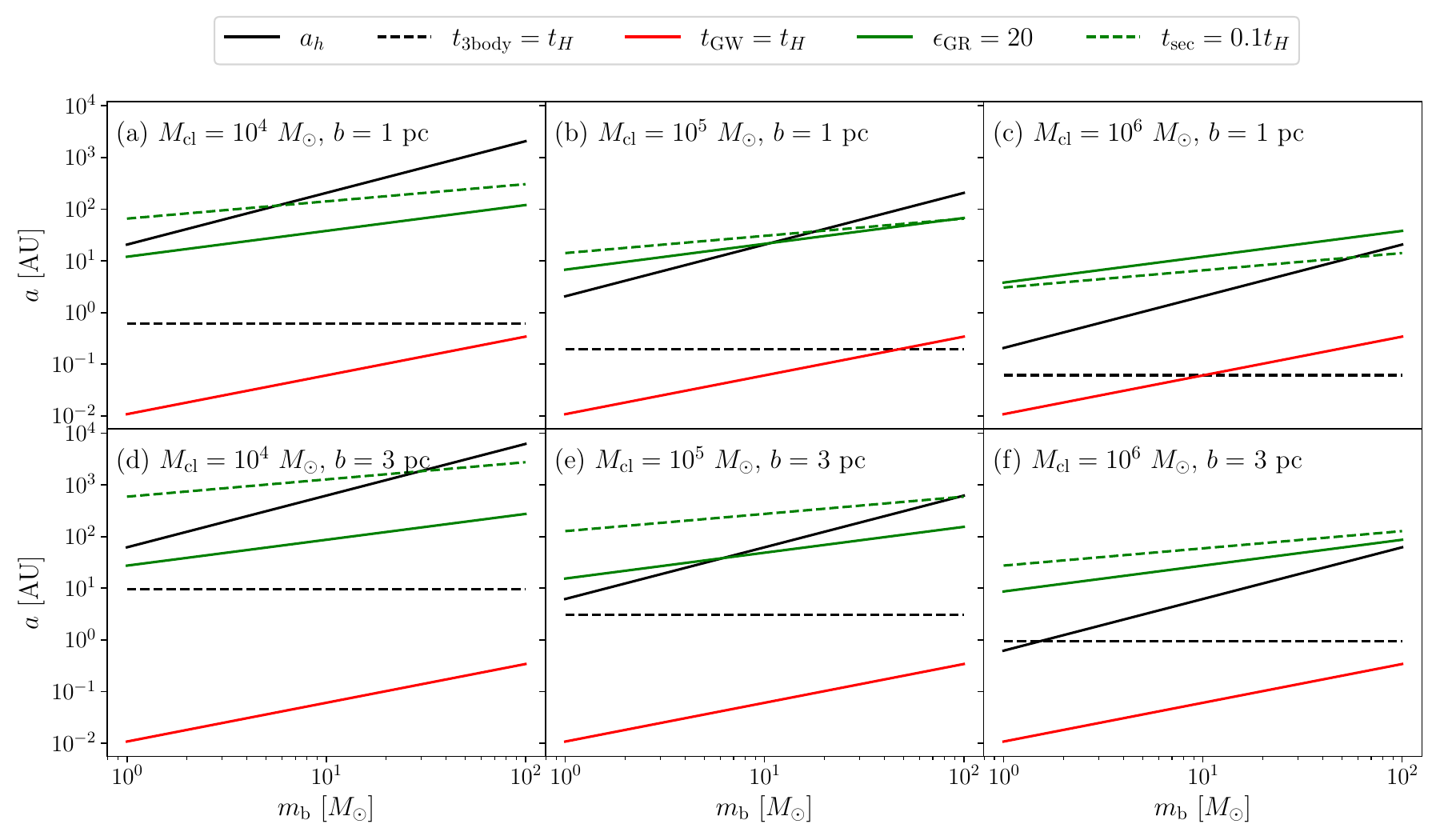}
 \vspace {-0.7cm}
\caption{
Characteristic values of the inner semimajor axis $a$ delineating different physical regimes for a binary evolving in a stellar cluster. Different curves correspond to the following: (black solid) hard/soft binary separation $a_h$, (black dashed) the semimajor axis $a_\mathrm{3body}$ at which 3-body hardening timescale $t_\mathrm{3body}$ becomes shorter than Hubble time, (red solid) $a=a_\mathrm{GW}$ at which the GW decay timescale becomes shorter than Hubble time, (green solid) $a=a_\omega$ at which GR precession suppresses cluster tides ($\epsilon_\mathrm{GR}=20$), (green dashed) $a=a_\mathrm{sec}$ at which the secular timescale is equal to 0.1 Hubble time. The cluster has characteristic radius $b=\SI{1}{pc}$ (top) or $\SI{3}{pc}$ (bottom) and total mass $M_\mathrm{cl}=10^4\,\msun$ (left), $10^5\,\msun$ (middle) or $10^6\,\msun$ (right). See \S\ref{section:relativeImportance} for additional details.
 }
\label{fig:timescales}
\end{figure*}


\subsection{Relative importance of different physical effects}
\label{section:relativeImportance}


Efficiency of the physical processes listed above depends on the binary semimajor axis $a$ and binary mass $\mb$. To illustrate this, in Fig.~\ref{fig:timescales} we show some characteristic values of the inner semi-major axes $a$ separating different physical regimes (explained below), as a function of the binary mass $\mb$. Our calculation assumes a cluster with a total mass $M_{\rm cl}$ and characteristic radius $b$. In all the estimates below we assume the radius of the binary outer orbit to be $\sim b$ (which essentially makes the exact form of the cluster potential irrelevant). Different panels in Fig.~\ref{fig:timescales} correspond to different values of $M_{\rm cl}$ and $b$. The meaning of the different lines is as follows:
\begin{enumerate}
\item Black solid: ``Hard binary separation'' $a_h=Gm_2/(4\sigma^2)$ \citep{quinlan} for a binary with component masses $m_2<m_1$ ($m_1+m_2=\mb$) and the local velocity dispersion $\sigma$ of cluster stars with mass $m_3\lesssim m_2$. This semi-major axis provides a separation between the 'hard' binaries that tend to harden (shrink their semi-major axes) in stellar encounters and 'soft' binaries, which are softened (driven to expand) by stellar flybys \citep{Heggie1975}. As demonstrated by \citet{quinlan}, binary hardening due to stellar ejections becomes efficient only at $a<a_h$, when the 3-body hardening rate $\de (a^{-1})/\de t$ becomes approximately constant. Thus, $a_h$ represents an important scale separating two distinct regimes of the possible binary semi-major axis evolution due to stellar encounters.
\item Black dashed: the semimajor axis $a_{\rm 3body}$ at which 
the hardening timescale \citep{quinlan}
\eq{
t_h \equiv \qty[a\dv{}{t}\qty(\frac{1}{a})]^{-1} = H^{-1}\frac{\sigma}{G\rho a},
\label{eq:th}
}
where $H\approx 15$ is a constant factor and $\rho$ is the local density of cluster stars, becomes equal to the Hubble time, 
$t_h = t_H$. With this definition, 3-body hardening is an important actor and should be taken into account when following binary evolution whenever $a_{\rm 3body}<a<a_h$. 
\item Red: the semimajor axis $a_{\rm GW}$ at which the merger timescale due to GW emission becomes equal to the Hubble time, $t_{\rm GW}=t_H$, assuming a circular binary. GW emission is efficient (on timescales $\sim t_H$) for circular binaries when $a<a_{\rm GW}$; for eccentric binaries, $a_{\rm GW}$ can be much higher (see \S\ref{section:GREffects} for more details).
\item Green dashed: the semimajor axis $a_{\rm sec}$ above which the period of secular oscillations of the orbital elements due to the cluster tides (see \S\ref{section:tidalEffects}) becomes shorter than $0.1t_H$; we choose this somewhat arbitrary value to allow for at least a few oscillations to occur in the Hubble time, letting GW emission to shrink the binary orbit during the high-$e$ episodes. With this logic, secular evolution driven by cluster tides is efficient when $a>a_{\rm sec}$. 
\item Green solid: the semimajor axis $a_\omega$ below which the binary pericentre precession due to GR effects becomes considerably faster than the cluster tide-driven precession. When this happens, the GR precession effectively suppresses eccentricity growth near its peak in the course of secular oscillations, severely reducing GW emission and semi-major axis evolution \citep{HR2021}. We calculate $a_\omega$ using the constraint $\epsilon_{\rm GR}=20$ on the dimensionless parameter $\epsilon_{\rm GR}$ characterizing the strength of the GR precession, see \S\ref{section:GREffects} for details (we also set $A=0.5GM_{\rm tot}/b^3$ in Eq.~\eqref{eq:epsilon_GR}). With this constraint in mind, cluster tides coupled with the GW emission would be efficient at shrinking the orbit of the binary when both $a>a_{\rm sec}$ (to enable secular oscillations in the first place) and $a>a_\omega$. 
\end{enumerate}

Careful examination of the Fig.~\ref{fig:timescales} allows us to make the following observations:
\begin{itemize}
\item Cluster tides are important only at sufficiently high semimajor axes, tens to hundreds of AU, especially for higher $\mb$. Only in densest (more massive and more compacts) clusters does $a_{\rm sec}$ decrease to several AU (making cluster tides more important) for $\mb\sim M_\odot$, typical for neutron star-neutron star (NS-NS) binaries, see panel (c). 
\item The 3-body hardening is more efficient than softening at $a\lesssim a_{\rm sec}$, when the cluster tide effects are usually not so important. However, for lower $M_{\rm cl}$ and smaller $b$ there is a small range of $a$ at high $\mb$ in which both hardening and cluster tides are important, i.e. where $a_{\rm sec}<a<a_h$. 
\item Hardening is much more efficient for heavier binaries, as they enter the ``hard regime'' earlier (since $a_h\propto m_2$), and once  $a\lesssim a_h$ the hardening timescale $t_h$ becomes independent of the binary mass, see equation (\ref{eq:th}). The semimajor axis range in which the 3-body hardening is important also shifts downwards in denser clusters; e.g. for $m_1=m_2=10M_\odot$ it is  $[0.1,1]$ AU for $M_{\rm cl}=10^6\,\msun$, $b=1$ pc (panel (c)) but $[10,1000]$ AU for $M_{\rm cl}=10^4\,\msun$, $b=3$ pc (panel (d)).
\item Unless the binary has very high eccentricity, $1-e\ll1$, GW emission is able to drive the binary to merger only at very low semimajor axes (0.01-0.1 AU) when the 3-body hardening is typically inefficient (except for the densest clusters, see panel (c)). Thus, the only two ways for the binary to achieve merger is to either (1) reach extreme eccentricity (via either cluster tides or 3-body interactions) or (2) have its semimajor axis substantially reduced as a result of a strong 3-body scattering event, such that the GW emission becomes important for shrinking its orbit. The results of both Monte-Carlo \citep{RodriguezLoeb} and population synthesis \citep{AntoniniGieles} codes confirm that heavier and/or denser clusters have higher BH merger rates. 
\end{itemize}

One important caveat to keep in mind is that in making this plot we neglected the stochastic component of the 3-body encounters and only took into account the average 3-body hardening rate. As we will show in \S\ref{section:results}, the real effect of random encounters is greatly enhanced by their stochastic nature, for which $a_h$ and $t_h$ are not very representative. Investigating the true role of stellar flybys on the binary orbital evolution will be the subject of Rasskazov \& Rafikov (in prep.).


\section{Ingredients of our numerical method}
\label{section:Physics}


Our numerical framework BESC is designed to follow the evolution of the outer and inner orbits of an individual binary in a stellar cluster, provided a set of binary and cluster parameters and initial conditions --- masses of the components $m_1$ and $m_2$, semimajor axis $a$, eccentricity $e$, inclination $i$, argument of periapsis $\omega$ and longitude of ascending node $\Omega$, as well as the initial position and velocity of the binary within the cluster. BESC does not handle the production (although it accounts for their destruction, see \S\ref{section:examples}) of binaries --- they can be either primordial or formed as a result of dynamical interactions in a cluster. Note that all Keplerian elements of the inner orbit are defined with respect to the reference plane and direction inside the cluster, which are specified in the beginning of the calculation and remain fixed throughout the evolution. 

We also specify the density profile of the cluster $\rho(r)$ as a function of distance from its centre $r$, which allows us to obtain the potential $\Phi(r)$ by solving the Poisson equation. In this work we focus on spherically-symmetric clusters; however, our treatment of cluster tides can be trivially extended to axisymmetric \citep{paper1} and fully triaxial \citep{BubPetrovich} systems. Our framework allows for a possibility of a mass spectrum of cluster stars, specified via the mass distribution function $f(m)$ (normalized such that $\int f(m)\dd m=1$), but in the examples shown in this paper we will focus on a single-mass case for simplicity.

In this section we provide the details of implementation of each individual physical process accounted for by our framework (\S\ref{section:encounters}-\ref{section:GREffects}), and then describe their synthesis into a framework for following the evolution of both the outer and the inner (\S\ref{section:EvolutionOfTheOuterOrbit}) orbits of the binary.


\subsection{Random encounters with the cluster stars}
\label{section:encounters}


Encounters with cluster stars play a very important role in binary evolution. Unlike cluster tides, they can change not only the eccentricity and inclination but also the semi-major axis of the binary. Moreover, they affect not only the inner but also the outer orbit of the binary, as well as its overall composition and fate (\S\ref{section:outcomes}). We will now provide a description of how we compute the rate of stellar encounters (\S\ref{section:encounterRate}), model each individual encounter (\S\ref{section:3bodymodeling}), and calculate the changes of the binary inner (\S\S\ref{section:changes_dist}, \ref{section:changes_close}) and outer (\S\ref{section:DF}) orbits resulting from a stellar flyby.


\subsubsection{Encounter rate}
\label{section:encounterRate}

To determine when an encounter with a cluster star happens, we use the result of \citet{HamersTremaine} for the mean rate of stellar encounters with pericenter distances (relative to the binary CoM) below some chosen\footnote{In our notation $Q_\mathrm{max}$ coincides with the radius of encounter sphere $R_\mathrm{enc}$ used in \citet{HamersTremaine}.} $Q_\mathrm{max}$:   
\eq{
\label{eq:encounterRate}
\rate = 2\sqrt{2\pi}n\sigma_\mathrm{rel}Q_\mathrm{max}^2 \int \dd m_3 f(m_3) \qty[1+\frac{G(\mb+m_3)}{Q_\mathrm{max}\sigma_\mathrm{rel}^2}],
}
where $n$ is the stellar number density, $\mb$ and $m_3$ are the binary and the perturber masses, and $f$ is the stellar mass distribution in the cluster; the factor in brackets accounts for gravitational focusing. Also, $\sigma_\mathrm{rel}=\sqrt{\sigma^2+V^2}$ is an approximation for the relative velocity dispersion, where $\sigma$ is the cluster velocity dispersion, $V$ is the (instantaneous) binary CoM velocity, and for simplicity an isotropic velocity distribution in the cluster frame is assumed. The isotropic velocity dispersion profile $\sigma(r)$ is determined from the Jeans equation: 
\eq{
\dv{(\rho\sigma^2)}{r} = -\rho\dv{\Phi}{r},
\label{eq:Jeans}
}
where $\Phi(r)$ is the cluster potential.

If $\rate$ is approximately constant along the outer orbit (which is the case when the stellar density and velocity dispersion vary little along the outer orbit or when the mean time interval between encounters $\rate^{-1}$ is short compared to the period of the outer orbit), we can pick the time interval until the next 3-body interaction assuming the probability $p(t>\tau)$ for the time between the encounters $t$ to exceed some $\tau$ to be $p(t>\tau)=\exp(-\rate\tau)$, as in \citet{HamersTremaine}. However, in many cases, we have to account for the variation of $\rate$ along the trajectory, and we do this by adopting the following probability density distribution:
\subeq{
\label{eq:probability}
\dd{p}(t>\tau) &= \exp\qty(-\int_0^\tau\rate(t')\dd{t'})\rate(t)\dd{t} \equiv \exp(-x(\tau))\dd{x},\\
x(t) &\equiv \int_0^t\rate(t')\dd{t'},
}
reducing to the \citet{HamersTremaine} assumption for constant $\rate$. Here $\dd{p}$ is the probability that the encounter happens between $t$ and $t+\dd{t}$ and $x$ is the dimensionless auxiliary variable. Therefore, after each encounter, we determine the time $t_\mathrm{nxt}$ until the next encounter in the following way: 
\begin{enumerate} 
\item Sample the value of $x=x_\mathrm{nxt}$ from the exponential distribution, see equation (\ref{eq:probability}). 
\item Integrate in time the outer orbit, the inner orbit (as described in \S\ref{section:EvolutionOfTheOuterOrbit}), and the equation 
\eq{\label{eq:dxdt}
\dv{x}{t} = \rate(t),
}
where $x(0)=0$ and $\rate(t)$ is computed via the equation (\ref{eq:encounterRate}) at the current location of the binary CoM.
\item The moment of time when $x=x_\mathrm{nxt}$ is $t=t_\mathrm{nxt}$ and we assume that the next encounter has taken place.
\end{enumerate}

This procedure allows us to naturally account for the spatial inhomogeneities of $n$ and $\sigma$ inside the cluster, which are sampled by the binary along its outer orbit.


\subsubsection{Modeling of encounters}
\label{section:3bodymodeling}

Once an encounter happens, we model it in the following way.

First, we randomly sample its initial parameters:
\begin{enumerate}
\item The perturber mass $m_3$ from the distribution function $f(m_3)$. In this paper we ignore the stellar mass spectrum and consider all perturbers to have the same mass. More realistically, $m_3$ could be drawn from e.g the Salpeter distribution $f\propto m_3^{-2.35}$ \citep{salpeter} modified to account for the finite lifetime of stars \citep[as was done in][]{HamersTremaine}.
\item The perturber initial velocity ${\bf v}_\mathrm{rel}$ (at infinity). To get it we first sample a velocity ${\bf v}$ in the cluster frame from the local (isotropic) velocity distribution $f(v)$, and then subtract the binary CoM velocity ${\bf V}$ from it. The local velocity distribution is considered to be Maxwellian,
\eq{\label{eq:maxwell}
f(v)\dd{v} = \frac{4\pi v^2\dd{v}}{(2\pi)^{3/2}\sigma^3} e^{-v^2/2\sigma^2},
}
where $\sigma$ is the local velocity dispersion determined from the equation (\ref{eq:Jeans}). The  relative encounter velocity is then asymmetrically distributed, giving rise to the dynamical friction, which is discussed in more detail in \S\ref{section:DF}. 
\item The impact parameter $p$ of the perturbing star (or any other passing object). For an adopted maximum encounter pericenter distance $Q_\mathrm{max}$ and $v_\mathrm{rel}$ we calculate the maximum impact parameter $p_\mathrm{max}$ and then randomly select $p$ such that $p^2$ is uniformly distributed between 0 and $p_\mathrm{max}^2$.
\item The (uniformly distributed) orientation of the perturber orbital plane relative to its approach velocity at infinity (determined earlier).
\item The initial mean anomaly of the binary (distributed uniformly).
\end{enumerate}

In this work we adopt $Q_\mathrm{max}=50a$ for the maximum pericenter distance of the perturbing star, see Fig. \ref{fig:hybrid_integration}. Appendix \ref{appendix} provides justification for this particular value of $Q_\mathrm{max}$ as well as for our choices of other BESC parameters discussed below --- $Q_\mathrm{hyb}$, $r_\mathrm{3body}$ and $r_\mathrm{max}$. Our decision to include flybys as distant as $50a$ is motivated by the fact that the cumulative effect of distant encounters can have an important effect on the binary eccentricity, especially when $e$ is high, and, as a consequence, on the overall binary merger rate \citep{distantEncountersEffect}. However, that also means we have to process a very high number of flybys, scaling roughly as $\propto Q_\mathrm{max}^2$ and increasing as $a$ grows. Luckily, the most distant encounters have a very small effect on the binary (individually), and in their course the orbit of the perturber does not deviate much from the hyperbolic trajectory. Because of that, for the flybys above a certain value of $Q=Q_\mathrm{hyb}$ it is possible to reduce the computational cost by using some approximations instead of a full 3-body integration, as we describe next.


\subsubsection{Changes of orbital elements: distant encounters}
\label{section:changes_dist}

\begin{figure}
\centering
\includegraphics[width=0.49\textwidth]{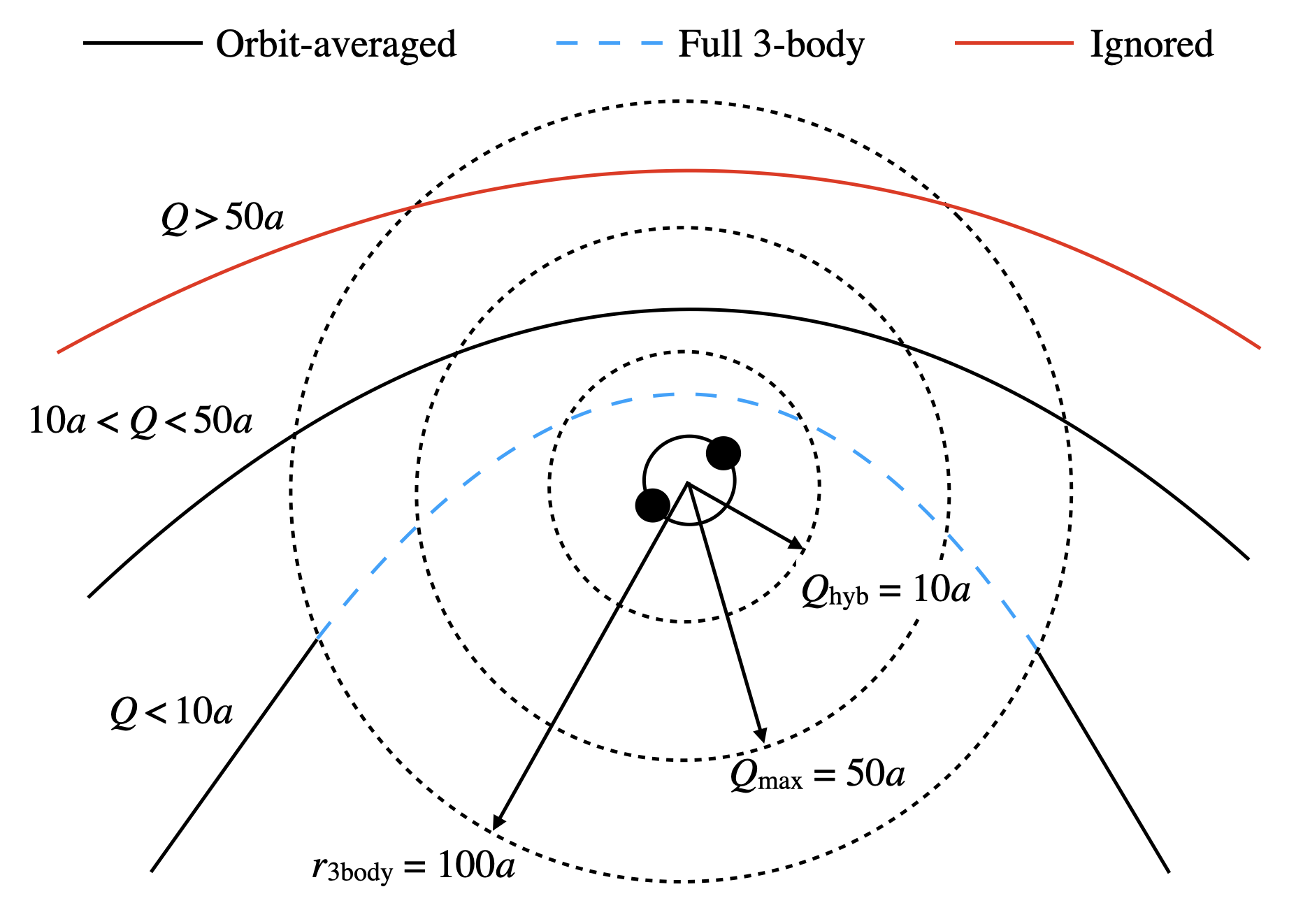}
\caption{
Schematic illustration of the different regimes of encounter integration used in BESC. We account for all flybys with pericentre distances of the third body $Q<Q_\mathrm{max}=50a$, but ignore the ones with $Q>Q_\mathrm{max}$ (red trajectory). When $Q>Q_\mathrm{hyb}=10a$, we use the semi-analytical orbit-averaged approximation of \citet{distantEncounters2} to integrate the encounter (black trajectory). When $Q<Q_\mathrm{hyb}$ (solid-to-dashed curve), we still use that approximation whenever the third body is more than $r_\mathrm{3body}=100a$ away from the binary centre of mass (black solid segment), but switch to the full 3-body integration whenever the incoming star approaches closer than $r_\mathrm{3body}$ (blue dashed segment). See \S\ref{section:3bodymodeling}-\ref{section:changes_close} for more details. 
}
\label{fig:hybrid_integration}
\end{figure}

Keeping track of the binary orbital elements is extremely important for enabling GW-assisted mergers of compact binaries, especially when the binary can be driven to high eccentricity. For that reason, we approach calculation of the binary inner orbital element changes with great care, in particular, making sure that we can accurately compute binary evolution at high eccentricities. Since at $e\to 1$ the angular momentum of the binary becomes very small, even the weak kicks experienced by the binary during its encounters with rather distant perturbers can have a significant effect on e.g. the binary eccentricity \citep{distantEncounters,distantEncounters2}. Given the large number of such distant encounters, the direct 3-body integration of each encounter is numerically prohibitive, so we used a different method based on the work of \citet{distantEncounters2}.

For distant perturbers, which we define as those with $Q>10a$, we use the orbit-averaged (OA) approximation from \citet[]{distantEncounters2}\footnote{Called `singly-averaged' (SA) in that paper. We do not use that term here to avoid confusion with the singly-averaged treatment of cluster tides, see \S\ref{section:SAvsDA}.} to calculate the changes of the orbital elements. This semi-analytic approximation is based on the evolution equations (Eqs. (2) of that work) for the eccentricity and (dimensionless) angular momentum vectors ${\bf e}$ and ${\bf j}$, accurate to octupole order (i.e. the perturbation on the binary from the third body is expanded in the series of $(a/r)^n$ up to $n=3$, where $r$ is the distance between the binary and the perturber) and assuming averaging over the binary orbital phase (i.e. over the inner orbit of the binary). Assuming also that the orbit of the perturbing star relative to the binary CoM is fixed (i.e. a Keplerian hyperbolic trajectory, considering the binary potential to be that of a point mass $\mb$), these evolutionary equations for ${\bf e}$ and ${\bf j}$ are then integrated numerically over the entirety of the perturber's orbit. This procedure was implemented by \citet{distantEncounters2} as a {\sc Python} script\footnote{https://github.com/hamers/flybys}, which we use in our calculations. 

In our case, we approximate this calculation by  integrating over the perturber's hyperbolic orbit from the moment it enters the sphere of radius $r_\mathrm{max}=10^4a$ (not shown in Fig. \ref{fig:hybrid_integration} for simplicity) to the moment it leaves that sphere. This implementation of the OA approximation is tested against the direct 3-body integrations in Appendix~\ref{appendix}, where we show, in particular, that adopting $r_\mathrm{max}=10^4a$ provides an accuracy sufficient for our purposes. At the same time, in most cases the computational expense of numerical integration over this radial range is acceptable, even for large number of distant encounters (but see \S\ref{section:examples}).

In principle, one may adopt other approaches to computing the changes of the binary orbital elements. For example,  \citet{distantEncounters} considered an approximate analytical solution of the OA equations in the `first-order approximation' -- when equations are integrated by assuming that all binary parameters stay constant in the course of an encounter; additionally these formulae assume a potential truncation at the quadrupole order. The first order approximation does not work well when the binary eccentricity is very high and the angular momentum $|{\bf j}|$ of the binary is low, so even a small perturbation can easily change $|{\bf j}|$ significantly during the encounter. Since we are often interested in the case of extremely eccentric binaries, that approximation is not suitable for us. 

This issue can be circumvented by switching to the `second-order' \citep{distantEncounters} or 'third-order' \citep{distantEncounters2} approximations that take into account the variation of the binary orbital parameters in the course of an encounter. However, that approach is not practically applicable to the octupole approximation for the potential (which needs to be resorted to for moderately close encounters), as the number of terms in the ensuing analytical expression exceeds 10,000 \citep[][Table 1]{distantEncounters2}. Such higher-order calculation is feasible for the quadrupole-order evolution equations considered in \citep[][Eqs. 5]{distantEncounters}, however, the quadrupole approximation is inaccurate in the case of an unequal-mass binary, which is also one of the possibilities we consider here. Because of that, we do not use any of the analytical approximations and just integrate the OA octupole equations instead. 

Averaging over the (fast) binary orbital motion, which is the backbone of the OA approximation, is only justified when the third body moves sufficiently slowly, i.e. when the mean motion of the binary is much faster than the angular speed of the perturber at periapsis of its hyperbolic orbit (so called secular encounter). As shown by \citet{distantEncounters}, that condition always applies for hard binaries. However, as clear from Figure \ref{fig:timescales}, the binaries where the effects of cluster tides are important, which are of great interest for us, are soft ($a\gtrsim a_h$) in most cases. For such binaries, one can show using the results of \citet[][see their Eqs. 1-3]{distantEncounters} that the OA approximation is valid only when 
\eq{
\label{eq:Rless1}
\qty(\frac{a}{Q})^2\frac{\sigma^2 a}{G \mb} \ll 1,~~~~~~\mbox{or}~~~~~~
\frac{Q}{a}\gtrsim \sqrt{\frac{a}{a_h}}
}
(for $m_3\lesssim \mb$, $m_1\sim m_2$). As will be discussed later, in our calculations we typically consider the binary to be ionized (disrupted) and stop the simulation when $a$ exceeds $10^3$ AU.
For a typical initial configuration considered in this paper ($m_1=m_2=10\msun$, $M_{\rm tot}\sim10^6\msun$, $b\sim\SI{1}{pc}$) the hard binary separation $a_h\sim\SI{10}{AU}$. Therefore, $a/a_h\lesssim 100$ and for $Q>10a$ --- the condition we adopted for considering an encounter as distant and for adopting the OA approximation --- the secular constraint \eqref{eq:Rless1} is fulfilled.


\subsubsection{Changes of orbital elements: close encounters}
\label{section:changes_close}

Some of the simplifying assumptions listed above (e.g. that of a secular encounter, purely Keplerian hyperbolic trajectory of the perturber, etc.) get gradually violated as the pericenter distance of an encounter becomes smaller than $Q_\mathrm{hyb}=10a$. In that case, switching to a full 3-body integrations, as has been done by many in the past \citep{HamersTremaine,cmc}, becomes inevitable.

To treat such situations in a computationally efficient way, we developed a `hybrid' approach: we still use the OA approximation when the distance between the binary and the third body exceeds $r_{\rm 3body}=100a$, switching to a precise 3-body integration only when the third body approaches the binary closer than $r_{\rm 3body}$. This scheme is illustrated in Fig.~\ref{fig:hybrid_integration}. The 3-body integrations are carried out using ARCHAIN, an implementation of algorithmic regularization developed specifically to treat small-$N$ systems \citep{archain}. During some particularly chaotic encounters, this switching between 3-body and orbit-averaged can happen multiple times as the perturber leaves and returns to $r_\mathrm{3body}$ sphere, sometimes with an exchange interaction between the binary and the third body (when the third body becomes bound to one of the binary components, and the other one is ejected).

While switching from the OA approximation to a full 3-body integration at  $r_{\rm 3body}$, we must provide the 3-body integrator with the initial condition for the mean anomaly of the binary, which is averaged over in the OA approximation. This initial phase of the binary is chosen randomly every time. The numerical errors resulting from the `hybrid' approximation and from choosing randomly the initial binary phase during the switchovers are discussed in Appendix~\ref{appendix}. To summarize, we find this approach to provide a sufficient accuracy for our purposes.

\begin{figure}
\includegraphics[width=0.49\textwidth]{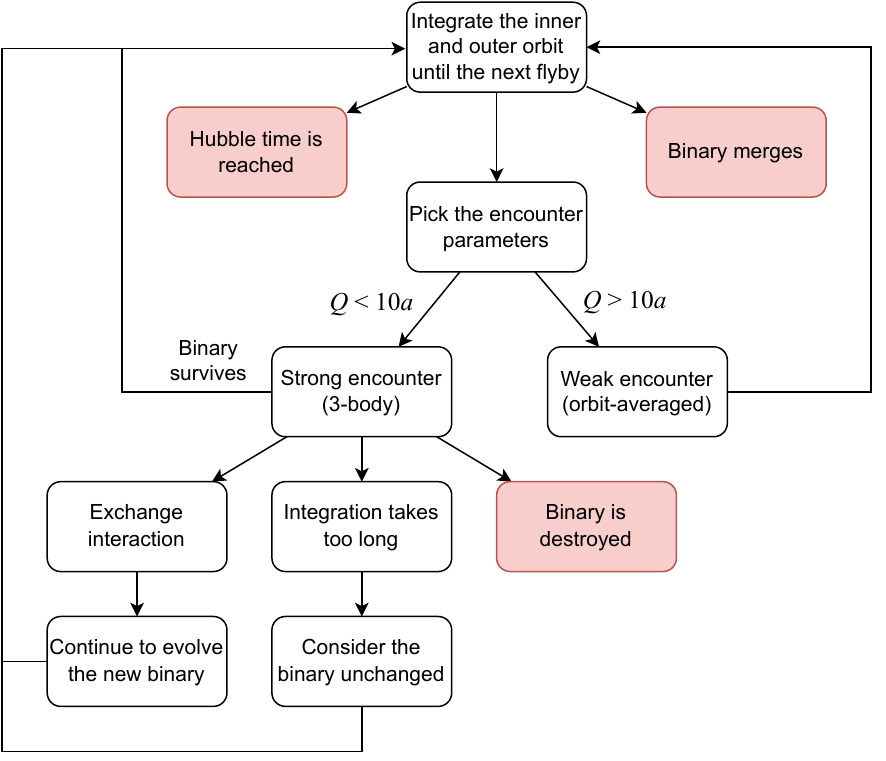}
\caption{
The flowchart illustrating the operational logic and major modules of BESC, as well as the possible evolutionary outcomes. See \S\ref{section:outcomes},\ref{section:EvolutionOfTheOuterOrbit} for details. }
\label{fig:flowchart}
\end{figure}

\subsubsection{Encounter outcomes}
\label{section:outcomes}

We stop the encounter calculation when one of the following outcomes takes place:
\begin{enumerate}
\item Two bodies are gravitationally bound to each other and form a binary, while the third one is at least $r_{\rm 3,max}$ away and is gravitationally unbound from the binary.
\item All three bodies become unbound from each other and are at a distance of at least $r_{\rm 3body}$ from each other -- in that case, we consider the binary to be destroyed (ionized).
\item The number of times when we switch to precise 3-body integration routine in our hybrid procedure (see \S\ref{section:changes_dist}) exceeds 20. In those highly chaotic cases (which constitute less than $10^{-5}$ of all encounters) we simply ignore the effect of that flyby.
\end{enumerate}

The flowchart illustrating these outcomes as part of our numerical framework is shown in Fig.~\ref{fig:flowchart}.

\subsubsection{Changes of the outer orbit and dynamical friction}
\label{section:DF}

Encounters with the cluster stars change not only the inner but also the outer orbit of the binary. After every encounter, we calculate the change in the binary CoM velocity, which modifies its outer orbit around the cluster. During the distant encounters ($Q>Q_\mathrm{hyb}=10a$) the binary is assumed to act on the incoming star as a point mass $\mb$ (see \S\ref{section:changes_dist}), thus, the change of the binary CoM motion is given by the outcome of a two-body encounter between the masses $\mb$ and $m_3$. For closer encounters ($Q<Q_\mathrm{hyb}$) the change of the outer orbit depends on the outcome of the 3-body integration which activates as a part of our hybrid method  (see \S\ref{section:changes_close}). 

At the same time, we assume encounters to be {\it local}, i.e. after an encounter the binary re-emerges at the same point in space inside the cluster. New CoM velocity after each encounter is used as an initial condition for integration of the post-encounter outer orbit of the binary, see \S\ref{section:EvolutionOfTheOuterOrbit}. 

Given the randomness of the encounter parameters (see \S\ref{section:3bodymodeling}), the outcome of each individual encounter is stochastic and leads to dynamical (non-resonant) relaxation. However, because the binary sees a greater flux of cluster stars arriving from the direction of its orbital motion, a large number of encounters leads to a net effect --- dynamical friction (DF) --- which causes the outer orbit of the binary to gradually decay towards the cluster centre (see \S\ref{section:examples}). This effect is directly captured by our treatment of encounters in BESC, without any semi-analytical modeling \citep[cf.][]{cmc}. 

At the same time, our constraint $Q<Q_\mathrm{max}=50a$ for an encounter to be considered may cause us to underestimate the magnitude of the DF. This is because DF is contributed not only by close but also by very distant encounters, with impact parameters comparable to the size of the cluster $b$. By lowering the maximum impact parameter to $\sim Q_\mathrm{max}$ we are effectively reducing the Coulomb logarithm in the expression for the DF \citep{BT} from $\ln[b\sigma^2/(G\mb)]\sim 8$ to $\ln[Q_\mathrm{max}\sigma^2/(G\mb)]\sim 5$, where we adopted typical values $b=1$ pc, $a=10^2$ AU, $\mb=20M_\odot$, and $\sigma=20$ km s$^{-1}$ (typical for $M_\mathrm{cl}=10^6M_\odot$). Thus, we expect our neglect of encounters with $Q>Q_\mathrm{max}$ to result in us underestimating the magnitude of DF by $\lesssim 50\%$, which we consider acceptable given the intrinsic uncertainties of the model.


\subsection{Secular evolution due to cluster tides}
\label{section:tidalEffects}


A distinct feature of BESC, present in only a handful of other studies \citep{paper3,BubPetrovich}, is the inclusion of the effects of cluster tides on the evolution of the inner orbit of the binary. 

Cluster tides arise because of spatial inhomogeneity of the cluster potential and cause binary orbital elements to vary in a secular fashion, without changing the binary semi-major axis. Their importance for the dynamics of relativistic (and other) binaries in stellar clusters has been pointed out by \citet{paper1,paper2,paper3}, who developed an analytical formalism for treating their effects; even earlier their significance was recognized for the dynamics of the Oort Cloud comets \citep{Heisler1986,Brasser2006}. \citet{paper2} explored the cluster tide-driven dynamics of a binary, showing that under certain circumstances a binary may be driven to very high eccentricities, similar to the Lidov-Kozai effect. As the intensity of the GW emission can rise dramatically during these high-eccentricity episodes, proper treatment of cluster tides is important for accurately following evolution of compact object binaries in a cluster.

Effects of cluster tides can be treated using two different levels of approximation. In a singly-averaged (SA) approximation, the evolution equations are averaged over the inner orbit of the binary only, and the local tidal field tensor (which serves as an input for the evolution) is sampled along the outer orbit of the binary in a cluster. In a doubly-averaged (DA) approach one also performs averaging of the tidal tensor over the outer orbit of the binary. As expected, the SA approach is more accurate and can reveal important effects which are hidden in the DA calculations, see \citet{paper5} and \S\ref{section:SAvsDA} of this work. This is why here we use the full SA evolution equations to model cluster tides. 

In BESC, we account for the tidal forces from the cluster potential using the numerical framework\footnote{https://github.com/mwbub/binary-evolution} based on Eqs.~(4)-(5) from \citet{BubPetrovich}, formulated in the SA approximation. In this framework, the outer orbit of the binary is first evolved using the \texttt{galpy} package \citep{galpy}. Then we calculate the tidal tensor from the cluster potential at $n$ points in this orbit, so that we could interpolate it later. We have established that 30 point per outer period is enough for almost all practical applications (apart from some rare cases where the eccentricity reaches extreme values, see \S\ref{section:SAvsDA}). Finally, the differential equations for the evolution of the inner binary parameters are solved, with the tidal tensor interpolated in space at every timestep as mentioned above.


\subsection{General relativistic effects}
\label{section:GREffects}


GR effects are the final key ingredients of our modeling framework. Emission of the GW is what causes the binary semi-major axis $a$ to shrink and ultimately drives the binary to merge. GR precession changes only the binary periastron, without affecting other orbital elements, but it still plays a key role in the dynamics.

We include the GR effects by adding the following terms to the evolution equations for the semimajor axis $a$ and eccentricity $e$ of the inner orbit and the argument of periapsis $\omega$ \citep{Peters1964}:
\subeq{
\qty(\dv{a}{t})_{\rm GR} &= -\frac{64}{5}\frac{q}{(1+q)^2}\frac{G^3(m_1+m_2)^3}{c^5a^3}\nonumber\\
&\times\frac{1+(73/24)e^2+(37/96)e^4}{(1-e^2)^{7/2}},
\label{eq:aGR}\\
\qty(\dv{e}{t})_{\rm GR} &= -\frac{304}{15}\frac{q}{(1+q)^2}\frac{G^3(m_1+m_2)^3}{c^5a^4}\nonumber\\
&\times\frac{1+(121/34)e^2}{(1-e^2)^{5/2}},
\label{eq:eGR}\\
\qty(\dv{\omega}{t})_{\rm GR} &= \frac{3e(G(m_1+m_2))^{3/2}}{a^{5/2}c^2(1-e^2)}, 
\label{eq:omegaGR}
}
where $c$ is the speed of light and $q=m_2/m_1$ is the binary mass ratio. Contributions given by the equations (\ref{eq:eGR}), (\ref{eq:omegaGR}) are added to the corresponding evolution equations for $e$ and $\omega$ from \S\ref{section:tidalEffects}. 

As discussed in \citet{paper2,HR2021}, when the semimajor axis $a$ is low enough and/or the eccentricity approaches unity, pericentre precession due to the GR becomes so fast (faster than the rate of cluster tide-driven evolution) that it suppresses the eccentricity oscillations arising from the cluster tides. This has important physical implications, reducing the maximum value that $e$ could reach and lowering the intensity of the GW emission, which considerably slows down binary evolution towards merger. The role of GR precession in counteracting the effects of cluster tides 
is characterized by the dimensionless parameter
\eq{
\label{eq:epsilon_GR}
\epsilon_{\rm GR} \equiv \frac{24G^2(m_1+m_2)^2}{c^2Aa^4},
}
where $A$ is the parameter \citep[defined in ][]{paper1} characterizing the strength of the cluster tide in the DA approximation; $A$ is of the order of the (squared) dynamical frequency of the cluster. One can think of $\epsilon_{\rm GR}$ as the ratio of the GR precession rate (\ref{eq:omegaGR}) for a circular binary to the rate of secular evolution due to cluster tides $t_\mathrm{sec}^{-1}\sim A P_\mathrm{b}$ ($P_\mathrm{b}$ is the binary period), down to factors of order unity. As demonstrated by \citet{HR2021}, for $\epsilon_{\rm GR}\gtrsim 20$ GR precession effectively suppresses the effects of cluster tides on binary evolution. 

Fast GR precession at low $a$ leads to certain numerical issues: in this regime the GR precession can be so fast that accounting for it in $\omega$ evolution numerically (necessary since the cluster tide contribution depends on $\omega$) requires a very small timestep, grinding the calculation to a halt. At the same time, in this regime the cluster tides are no longer important and accounting for them is not really necessary. 

For that reason we have adopted the following strategy: if at some point in the calculation (usually, after a close encounter that strongly reduces $a$) we find that 
\begin{align} 
\epsilon_{\rm GR}>20,
\label{eq:GRthreshold}
\end{align}
then we (1) stop calculating the contribution of cluster tides to the evolution of binary orbital elements and (2) evolve the binary apsidal angle analytically, using $\omega(t)=(\dd\omega/\dd t)_\mathrm{GR}t+\omega_0$ ($\omega_0$ is the initial value of $\omega$). A threshold value of 20 for $\epsilon_{\rm GR}$ was motivated by Fig. 9 of \citet{HR2021} and serves as a good compromise between the accuracy and numerical efficiency of the computation. This procedure allows us to avoid the aforementioned bottleneck due to the fast GR precession, dramatically speeding up the calculation. If at some later time the binary semi-major axis is increased (as a result of a  close encounter) and $\epsilon_{\rm GR}$ drops below 20, we restart the numerical calculation of the orbital evolution accounting for the cluster tides.

To calculate $\epsilon_{\rm GR}$, we average $A$ over the outer orbit \citep{paper1} until the next encounter. In many cases, the corresponding segment of the outer orbit covers only a few orbital periods around the cluster centre and $A$ may not have fully converged to its long-term average. That method, however, is good enough for our purposes, and quick enough to not slow down the computations. The $\epsilon_{\rm GR}(t)$ plots shown later in Figs.~\ref{fig:merged}--\ref{fig:tidalDominated} do not use these values of $A$ but rather the ones obtained using Eqs. (39) and (D7) from \citet{paper1}. We have checked that those agree on average with the values used in our numerical evolution.


\subsection{Summary: evolution of the outer and inner orbits}
\label{section:EvolutionOfTheOuterOrbit}


The three physical processes described above are combined together in a unified framework for integrating the evolution of the outer and inner orbits of the binary, which conceptually operates as follows (see Figure \ref{fig:flowchart} for an illustration of various steps).

\begin{enumerate}
\item After each encounter with a cluster star we sample the random variable $x_\mathrm{nxt}$, which eventually determines the time when the next encounter occurs (\S\ref{section:encounterRate}).
\item We then integrate the orbit of the binary CoM in the cluster potential using the \texttt{galpy} package \citep{galpy}. In this work, since we limit ourselves to the spherical potentials (a simplification that can be easily relaxed, see \S\ref{sec:future}) the turning points (inner $R_p$ and outer $R_a$) and the plane of the outer orbit are preserved between encounters.
\item We simultaneously evolve the orbital elements of the inner orbit of the binary due to the (smooth) effects of the cluster tides (\S\ref{section:tidalEffects}) and general relativity (\S\ref{section:GREffects}).
\item Also, knowing the encounter rate at any point of the outer orbit (see \S\ref{section:encounterRate}) we integrate the equation (\ref{eq:dxdt}) until $x$ reaches $x_\mathrm{nxt}$. 
\item At this point we declare the next encounter to take place, stop the integration of the inner and outer orbits and process the encounter as described in \S\ref{section:encounters}; the (impulsive) changes of the inner orbital elements (\S\S\ref{section:changes_dist}, \ref{section:changes_close}) and of the outer orbit (\S\ref{section:DF}) are used to update the binary parameters. 
\item We then determine the outcome of an encounter (\S\ref{section:outcomes}) and, if the binary survives, repeat the cycle. We consider the encounters to be instantaneous and local, thus we restart orbit integration from the time and point in space, at which the last encounter has taken place.
\end{enumerate} 

\begin{figure*}
\includegraphics[width=0.99\textwidth]{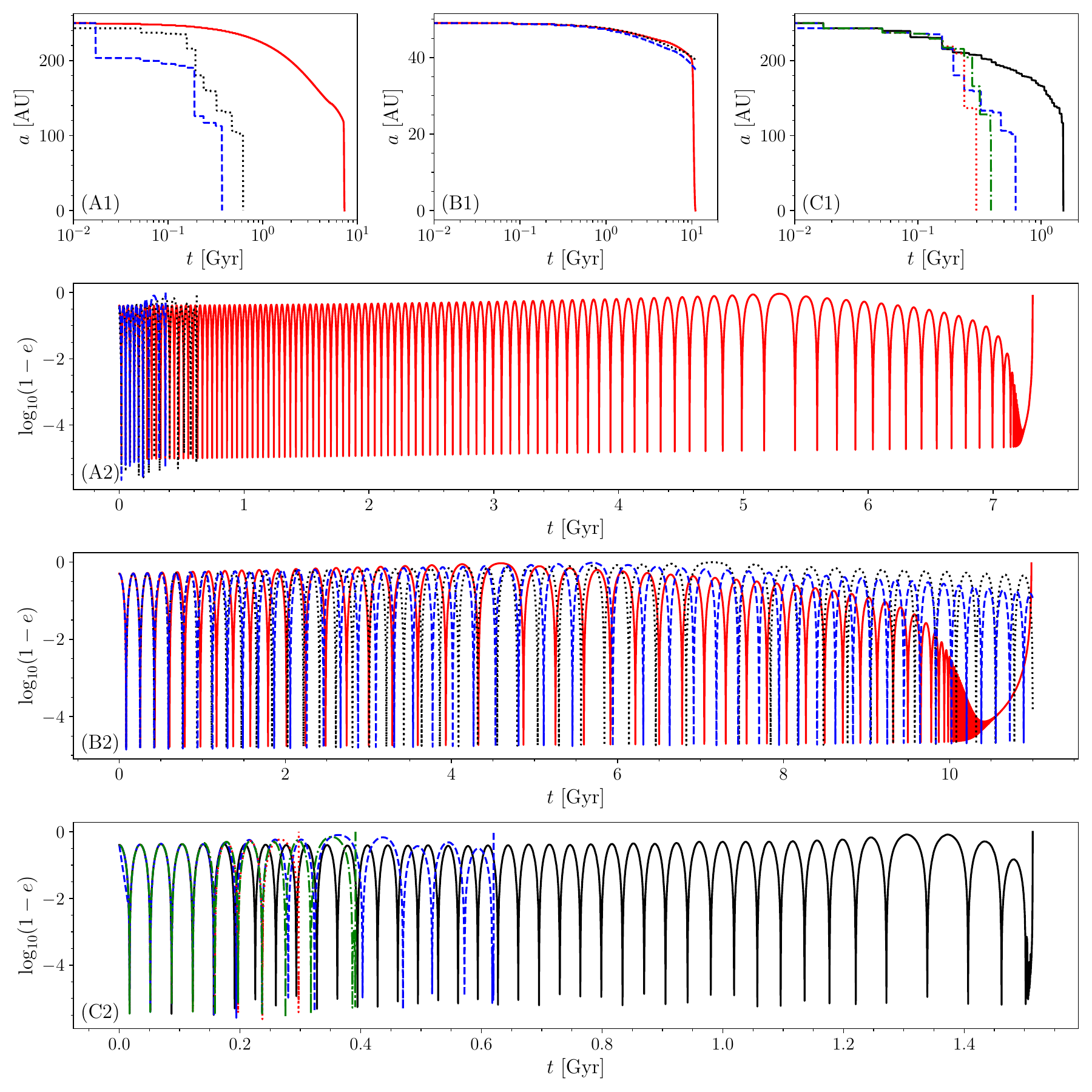}
\caption{
(A) Comparison of singly- and doubly-averaged approximations for binary evolution in the absence of encounters. Shown are the semimajor axis $a(t)$ (A1) and eccentricity $e(t)$ (A2) of a binary with the following initial conditions: $m_1=m_2=10\msun$, $a=\SI{250}{AU}$, $e=0.6$, $i=89.8^\circ$, $\omega=91^\circ$, $R_p=\SI{1.5}{pc}$, $R_a=\SI{1.7}{pc}$ in the Hernquist potential with $M_{\rm cl}=10^6\msun$, $b=\SI{1}{pc}$. Solid red shows doubly-averaged calculation from \citet[][example 2]{paper4}. Dotted black and dashed blue curves represent singly-averaged integrations using BESC with initial $\Omega=0$ and $\Omega=\pi/2$, respectively. (B) Same as (A) except for the different initial parameters: $m_1=m_2=1.4\msun$, $a=\SI{49}{AU}$, $e=0.5$, $i=89.9^\circ$ \citep[][example 3]{paper4}. (C) Illustration of the sensitivity of singly-averaged calculation to numerical parameters. Same initial conditions as in (A), with different lines corresponding to SA integration with the different number of interpolation points (per outer orbit) used to compute the potential derivatives when calculating the binary evolution: 30 (solid black), 100 (dashed blue), 300 (dotted red), 1000 (dot-dashed green). See \S\ref{section:SAvsDA} for more details on the chaotic nature of the singly-averaged evolution.}
\label{fig:comparison-with-da}
\end{figure*}

This way, BESC follows the binary evolution until it either merges or is destroyed, or until a Hubble time has passed at which point we stop the calculation.


\section{Results}
\label{section:results}


We now show the results of some tests of our framework, first focusing on the cluster tide-driven evolution and turning off the effects of encounters (\S\ref{section:SAvsDA}). This allows us to compare our framework against some known results and to demonstrate the importance of using the SA approximation for cluster tides. We then present some illustrative examples of binary evolution including both cluster tides and encounters with cluster stars (\S\ref{section:examples}).


\subsection{Evolution in the absence of encounters: comparison with the DA approximation}
\label{section:SAvsDA}


In Fig.~\ref{fig:comparison-with-da} we show some examples of binary evolution calculated using BESC with the stochastic stellar encounters completely turned off, allowing us to focus on the effect of cluster tides and to compare the performance of the DA and SA approximations (\S\ref{section:tidalEffects}). Top row shows the evolution of the semi-major axis $a$ while the bottom panels illustrate the behavior of $1-e$ for binaries that reach very high eccentricities as a result of the specially chosen initial conditions --- the initial binary inclination (relative to the plane of the outer orbit) is very close to $90^\circ$. We first focus on panels (A) and (B), in which we compare the evolutionary tracks of the binaries computed using the doubly-averaged \citep[red, reproduced from Figs. 4 and 5 of][which should be consulted for parameters of these calculations]{paper4} and singly-averaged (blue and black, computed using BESC) approximations with the same initial conditions, correspondingly.  

The DA track in Fig.~\ref{fig:comparison-with-da}A shows an initially librating binary in a weak GR regime (see \citet{paper4} for a detailed discussion of the different evolutionary phases), capable of reaching an extreme eccentricity $e_\mathrm{max}\approx1-10^{-5}$ in the course of cluster tide-driven secular oscillations. Its phase space trajectory eventually crosses the separatrix and becomes circulating (around $t\approx\SI{4.8}{Gyr}$), then the binary merges at $t\approx\SI{7.4}{Gyr}$. Fig.~\ref{fig:comparison-with-da}B shows another example from \citet{paper4}: a lower-mass binary with smaller initial semimajor axis $a$, but with the same outer orbit in the same potential. It also starts with a librating phase space trajectory (although in the moderate GR regime) and goes through the same phase space transformations as the one in the example A before merging (see \citet{paper4} for details). 

As to the SA results, they are initially similar to the DA tracks, but start to diverge from them after a few secular cycles. In addition to that, evolution in SA approximation sensitively depends on the initial longitude of ascending node $\Omega_0$ of the binary, which is the parameter distinguishing the black and blue tracks. The reason for this is that SA integration accounts for additional eccentricity fluctuations on the outer orbital period timescale, on top of the large-amplitude, smooth DA evolution of $e$ on a much longer secular timescale. These eccentricity fluctuations are typically negligible, but they can become very significant for $e\to 1$, as even a small change in $e$ can strongly affect the angular momentum of the binary and amplify its GW emission. 

In the example B those SA oscillations are small (peak eccentricity is almost the same for DA and SA tracks), and the DA-SA discrepancy increases gradually over many secular cycles, with both SA examples going through the same phase space transformations as DA, but with a delay (and as a result, merging much later than the DA track shows).

On the other hand, in the example A the eccentricity oscillations of the SA tracks are much larger, which is obvious from their deviations from the DA track at highest $e$. As a result, even though the timing of the first 6--7 eccentricity maxima is approximately the same, an abrupt divergence between the three evolutionary trajectories in both $e$ and $a$ happens soon after. As a consequence of the much stronger GW emission during the high-$e$ episodes on the SA tracks, both SA integrations in panels (A) result in an order of magnitude shorter binary merger times than predicted by the DA calculation.

Interestingly, even in the absence of GW emission (which we can easily switch off in BESC, not shown here) but with the GR precession still present, the SA trajectories diverge from the DA ones much faster in the example A than in the example B. This behavior is a manifestation of the so-called relativistic phase space diffusion (RPSD), which was studied by \citet{paper5}. It takes place when, in the course of its SA evolution, the binary can reach eccentricity so high that the time it spends around the peak of $e$ (over which GR precession drives a large-amplitude swing of the apsidal angle) becomes shorter than the outer orbital period of the binary. This causes sharp jumps of the (otherwise well conserved) secular integrals of motion of the binary (i.e. its perturbation Hamiltonian) and leads to a noticeable evolution of the period of secular oscillations (in the absence of both flybys and GW emission). 

\citet{paper5} have derived the following criterion for the SA eccentricity oscillations to be significant and, at the same time, for RPSD to occur (see their Eq. 37):
\eq{
|\cos i_0| &\lesssim 0.007 \times\qty(\frac{\mb}{2.8\msun})^{-1/2} \qty(\frac{a}{\SI{50}{AU}})^{3/2}  \nonumber\\
&\times \qty(\frac{M_{\rm cl}}{10^7\msun})^{1/2} \qty(\frac{R}{\SI{1}{pc}})^{-3/2},
}
i.e. that the binary inner orbital plane must be very close to orthogonal to the plane of the outer orbit (a necessary condition for reaching a very high $e$). One can easily check that for the initial parameters specified in the captions of Fig.~\ref{fig:comparison-with-da} the example A satisfies this condition, while the example B does not, which explains their different SA-DA behavior even in the absence of GW emission.

Since \citet{paper5} did not include GW emission when comparing DA and SA approximations for cluster tides, the examples shown in Fig.~\ref{fig:comparison-with-da} provide the first illustration of the coupling between the RPSD and the GW emission. Their synergy makes the binary merge much sooner, as even a small RPSD-related increase of $e$ around the eccentricity maximum can significantly increase the rate of the semimajor axis decay.  

Although only two SA trajectories (for two values of $\Omega_0$) are shown in Fig.~\ref{fig:comparison-with-da}A,B for every set of initial conditions, SA trajectories calculated for other initial $\Omega_0$ show similarly erratic divergence from the DA track and from each other. This illustrates the essentially chaotic behaviour of the SA oscillations, in a sense that even a tiny change in the input parameters of the calculation can lead to a significant change in the binary evolution. 

This statement is true not only for physical inputs (e.g. varying $\Omega_0$ in the examples A and B), but also for numerical parameters of the calculation, as we illustrate in Fig.~\ref{fig:comparison-with-da}C. There we show (for the same initial conditions as in the example A) the result of varying the number of points $n$ on the outer orbit that are used to approximate the derivatives of the gravitational potential when computing the tidal tensor for the SA calculation (see \S\ref{section:tidalEffects}). One can see that when $e$ periodically reaches extreme values, results of the calculation become very sensitive to the value of $n$, and the SA tracks show no obvious convergence even as we increase $n$ to $n=1000$. This means that the long-term binary evolution in the SA approximation is realistically hardly predictable, and that one can draw meaningful conclusions only upon examining the evolution of large samples of binaries with similar initial conditions. 

As we will see next, inclusion of encounters adds so much additional stochasticity to the binary evolution that this completely obviates the need to worry about the sensitivity to various input parameters. For practical applications --- calculating the evolution including encounters --- evaluating cluster potential at $n=30$ points turns out to be fully adequate.


\subsection{Binary evolution with all physical effects included: examples of different outcomes}
\label{section:examples}

\begin{figure}
\includegraphics[width=0.49\textwidth]{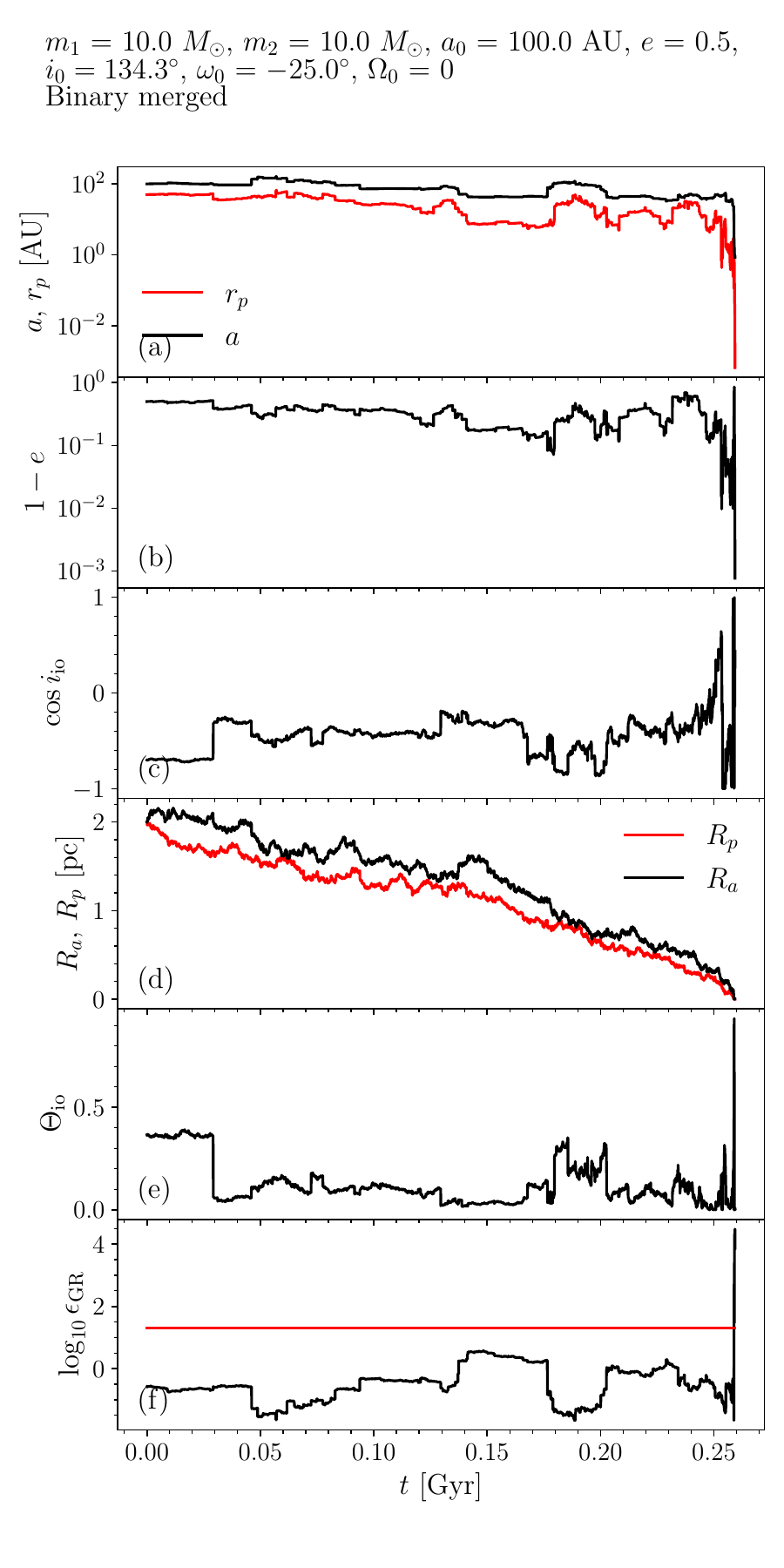}
\vspace{-1cm}
\caption{
Time evolution of various binary characteristics computed using BESC including all physical ingredients (cluster tides, encounters, GR), for an evolutionary track that results in a binary merging. Cluster has a Hernquist potential with $M_\mathrm{cl}=10^6\msun$ and radius $b=2$ pc, binary component masses are $m_1=m_2=10\msun$ and  perturbers have mass $m_3=1\msun$; see \S\ref{section:examples} and the legend at the top for other initial binary parameters. We show time evolution of the binary inner orbital characteristics: (a) periastron distance $r_p$ and semi-major axis $a$, (b) eccentricity $e$, (c) inclination $i_\mathrm{io}$, relative to the instantaneous plane of the outer orbit. Panel (d) shows the outer ($R_a$) and inner ($R_p$) radii of the outer orbit of the binary. Panel (e) illustrates the behavior of $\Theta_\mathrm{io} =(1-e^2)\cos^2{i_\mathrm{io}}$, relevant for cluster tides, while panel (f) shows $\epsilon_\mathrm{GR}(t)$ defined by the equation (\ref{eq:epsilon_GR}) and illustrating the importance of GR precession (red line is the constraint (\ref{eq:GRthreshold})). See \S\ref{sec:ex1} for details.
}
\label{fig:merged}
\end{figure}

We now illustrate the performance of BESC with all the physics (described in \S\ref{section:Physics}) fully incorporated, including stellar encounters. Figs. \ref{fig:merged}--\ref{fig:nokicks} show several examples of individual evolutionary tracks of BH-BH binaries illustrating a variety of possible outcomes: a merger (Fig.~\ref{fig:merged}), an impulsive binary disruption (ionization, Fig.~\ref{fig:destroyed}), an exchange interaction (Fig.~\ref{fig:exchange}), a diffusive disruption of a binary ($a$ exceeding $10^3$ AU, Fig.~\ref{fig:abandoned}), an ejection from the cluster (Fig.~\ref{fig:ejected}), a binary that survives for a Hubble time (Fig.~\ref{fig:nokicks}). Each of these cases is discussed in \S\ref{sec:ex1}-\ref{sec:ex6}.

Unless specified otherwise, we have chosen the following initial parameters for these evolutionary tracks:
\begin{itemize}
\item Spherical Hernquist potential of the cluster $\Phi(r)=-GM_\mathrm{cl}/(b+r)$ with $M_\mathrm{cl}=10^6\msun$ and $b=\SI{2}{pc}$. Velocity dispersion is assumed to be isotropic and given by the Eq. (10) from \citet{hernquist}.
\item Circular outer orbit  with initial radius $R=b=2$ pc.
\item All cluster stars have masses $m_3=1\msun$.
\item Binary component masses $m_1=m_2=10\msun$.
\item Initial inner semimajor axis $a=\SI{100}{AU}$.
\item Initial binary eccentricity $e=0.5$.
\end{itemize}
The initial parameters, which are different for the individual tracks that we present are as follows:
\begin{itemize}
\item Initial inclination $i_0$ is drawn from a uniform distribution of $\cos i_0$.
\item Apsidal longitude $\omega_0$ is uniformly distributed (although the longitude of ascending node is set to  $\Omega_0=0$).
\end{itemize}
This ensures that some differences between the evolutionary tracks exist from the start, however, the key reason for their subsequent divergence is the stochasticity due to stellar encounters.

Our choice of a (cusped) Hernquist modes may be more suitable for nuclear star clusters than for globular clusters with their cored density profiles. However, it allows us to illustrate some outcomes that are unlikely in the cored models, e.g. see \S\ref{sec:ex5}.

Our choice of the initial semimajor axis $a$ is motivated by the following considerations. The initial value of the GR parameter $\epsilon_{\rm GR}(0)=0.26$ satisfies the constraint (\ref{eq:GRthreshold}), i.e. cluster tides are not suppressed by the GR precession initially. However, given a very steep dependence of $\epsilon_{\rm GR}$ on $a$, $\epsilon_{\rm GR}\sim a^{-4}$, this would no longer be the case if $a$ were much smaller than 100 AU; the threshold (\ref{eq:GRthreshold}) is reached at (from Eq.~\ref{eq:epsilon_GR})
\eq{
a &= \SI{29}{AU} ~\qty(\frac{\epsilon_{\rm GR}}{20})^{-1/4} \qty(\frac{A_\ast}{0.5})^{-1/4} \qty(\frac{\mb}{20\msun})^{1/2} \nonumber\\
&\times \qty(\frac{M_{\rm cl}}{10^6\msun})^{-1/4} \qty(\frac{b}{\SI{2}{pc}})^{3/4}.
}
Thus, in order to explore the effects of the cluster tides we should start with $a\gtrsim 50$ AU.

On the other hand, the higher the semimajor axis, the less probable it is for the binary to eventually merge, since softer binaries harden due to encounters less efficiently. In addition to that, when dealing with encounters in BESC the encounter rate grows approximately as $\sim a^2$ (Eq.~\ref{eq:encounterRate}), and for very large $a$ the number of encounters that needs to be followed becomes so high that the computational cost of the calculation becomes prohibitive. To avoid that, we stop the simulation whenever the binary semimajor axis reaches $10^3$ AU. We have not observed any cases of a binary going above that limit but then shrinking its orbit and merging. 

As a result, starting evolution at $a=10^2$ AU, roughly at the hard/soft boundary, appears to be a reasonable compromise between the two aforementioned limiting factors.

\subsubsection{Example 1: binary merger}
\label{sec:ex1}

\begin{figure}
\includegraphics[width=0.49\textwidth]{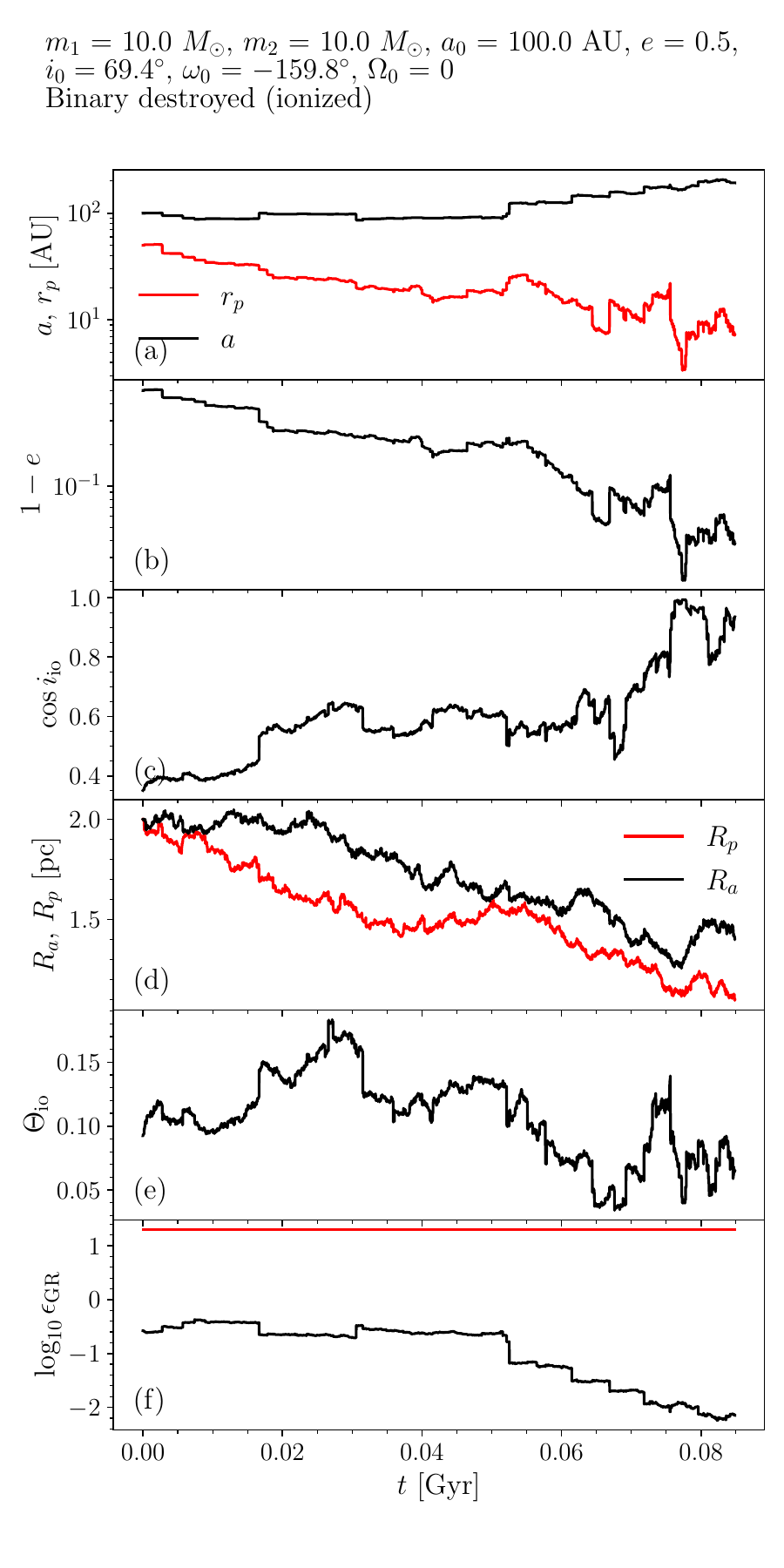}
\vspace{-1cm}
\caption{
Same as Fig.~\ref{fig:merged} except the binary is disrupted (ionized) by a 3-body encounter (see the legend at the top for the parameters of this evolutionary track). See \S\ref{sec:ex2} for a discussion. 
}
\label{fig:destroyed}
\end{figure}

\begin{figure}
\includegraphics[width=0.49\textwidth]{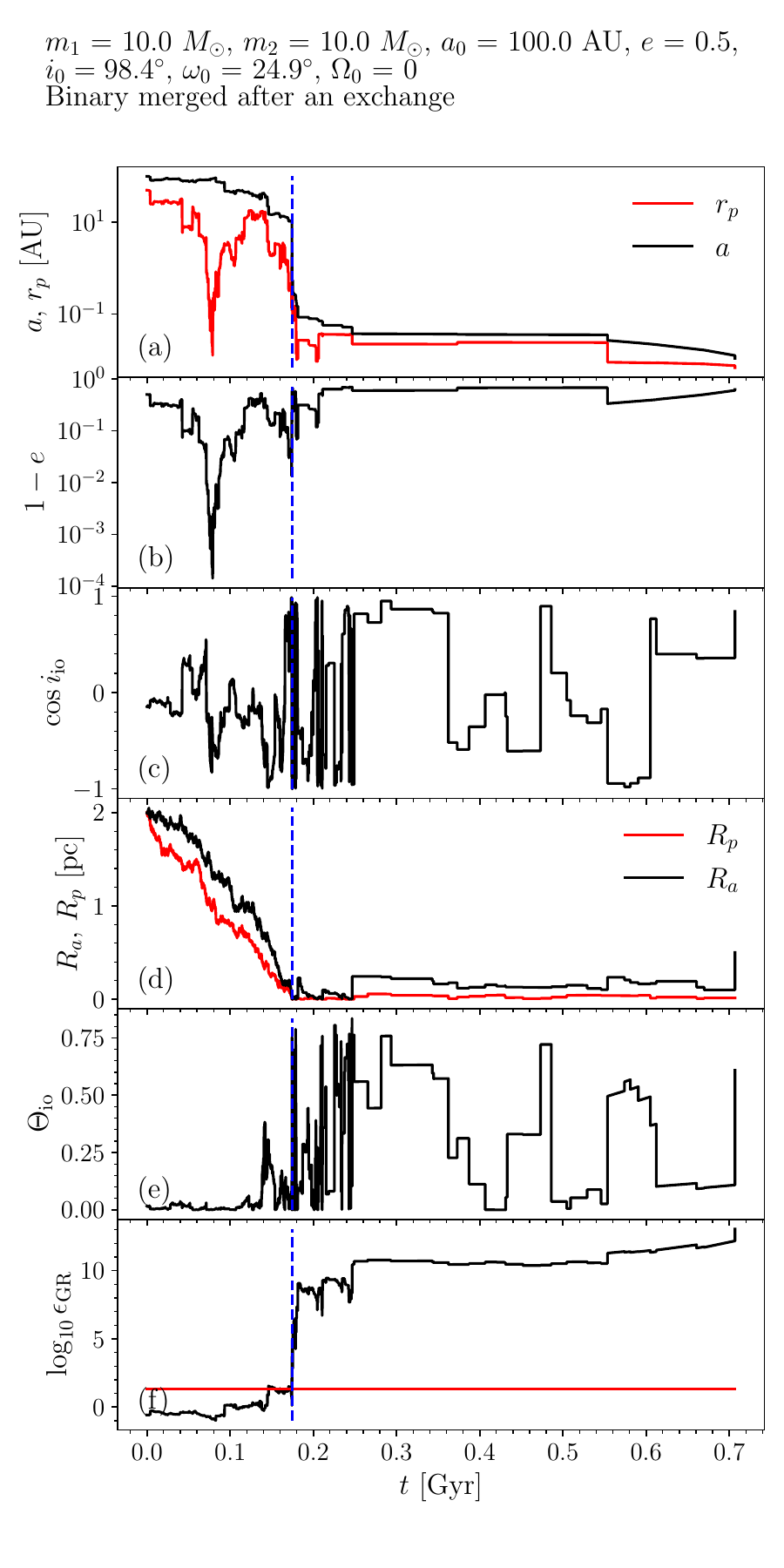}
\vspace{-0.8cm}
\caption{
Same as Fig.~\ref{fig:merged} except the binary experiences an exchange interaction at the moment marked by the blue dashed vertical line. See \S\ref{sec:ex3} for a discussion. }
\label{fig:exchange}
\end{figure}

Fig. \ref{fig:merged} illustrates the orbital evolution of a binary that ends up merging, and we use it to describe the presentation of binary characteristics in Figs. \ref{fig:merged}--\ref{fig:nokicks}, which all have the same structure. 

In panels (a)-(c) we show the 'raw' orbital parameters of the inner orbit --- (a) the semimajor axis $a$ and the periastron distance $r_\mathrm{p}$, which is indicative of the GW emission strength, (b) the deviation of the binary eccentricity $e$ from unity, and (c) the cosine of the inclination $i_\mathrm{io}$ of the inner orbit relative to the (instantaneous) plane of the outer orbit. Note that $i_\mathrm{io}$ is different from the inclination $i$ relative to the fixed cluster frame (set in the beginning of integration, when $i=i_\mathrm{io}$), since the plane of the outer orbit changes as a result of stellar encounters; see below for the significance of $i_\mathrm{io}$. In panel (d) we show the outer $R_a$ and inner $R_p$ turning points of the outer orbit of the binary. In panel (e) we show the evolution of an important dimensionless  parameter $\Theta_\mathrm{io}=(1-e^2)\cos^2 i_\mathrm{io}$, an essential integral of motion in both the DA \citep{paper2} and SA \citep{paper5} approximations. As shown in \citet{paper2}, $\Theta_\mathrm{io}\ll 1$ is a necessary condition for the cluster tides to be able to increase $e$ close to 1 for a binary that was not very eccentric initially, which means that initially $i_\mathrm{io}$ must be very close to $90^\circ$. Finally, panel (f) illustrates the behavior of $\epsilon_\mathrm{GR}$ (see Eq. \ref{eq:epsilon_GR}), which characterizes the importance of the GR precession in suppressing the effect of cluster tides. All these variables are shown as functions of time. 

One can see that the outer orbit of the binary gradually decays over time (panel (d)) as a result of dynamical friction, which is a consequence of multiple stellar encounters, see \S\ref{section:DF}. Encounters also result in rapid fluctuations of $R_p$ and $R_a$ but this diffusive effect is overwhelmed by the overall orbital decay. The outer orbit never becomes too radial as $R_p$ and $R_a$ remain similar to each other. Eventually, around $t=0.25$ Gyr the binary sinks to the centre of the cluster, where the encounter rate rapidly grows because of increased stellar density $\rho$ and velocity dispersion $\sigma$. This sinking time is in general agreement with the Chandrasekhar's formula with $\rho$ and $\sigma$ evaluated at the characteristic radius of the cluster $b$.

During this time, the inner orbit evolves as well (panels (a)-(c)). In this example it stays moderately eccentric through most of the evolution until eventually one of the stellar encounters, that are very frequent in the cluster center, sends $e$ to a very high value, $1-e\lesssim 10^{-3}$. With the periastron distance below $10^{-2}$ AU (panel (a)), the binary rapidly shrinks its orbit due to the GW emission and merges (the merger timescale for an $\mb=20M_\odot$, $q=1$ binary with $a=100$ AU and $r_p=0.01$ AU is around 4 Myr, see \citealt{Peters1964}). 

Panel (f) shows that until the very last moment $\epsilon_\mathrm{GR}$ stays below the critical value of 20. This implies that cluster tides are not suppressed by the GR precession and are being accounted for by BESC. Given that the initial inclination of the binary is not close to $90^\circ$, cluster tides acting alone are unable to increase $e$ to values close to unity. But they are still capable of changing $e$ by $\sim 0.1$ in the course of the evolution. Nevertheless, the final increase of $e$, which ultimately results in a merger, is due to stellar encounters, which are abundant in the cluster centre --- this manifests in the orbital parameters starting to change rapidly at the end of the binary lifetime. We can thus conclude that in this example it is the encounters and not the cluster tides that drive the binary to merger.

This can also be inferred from the behavior of $\Theta_\mathrm{io}$ (panel (e)) that becomes very small (while $e$ is not close to unity) only for a short period of time in the end of the evolution. This does not give cluster tides a chance to increase $e$ of this binary to values close to unity \citep{paper2}.

\subsubsection{Example 2: binary ionized impulsively}
\label{sec:ex2}

Fig. \ref{fig:destroyed} shows a binary that ends up being destroyed (ionized) by experiencing a close flyby that causes all three bodies to become unbound from each other. It starts from the same initial conditions as in the previous example (except for the initial orientation of the inner orbit), and the only reason for the different outcome is the stochastic nature of the flybys. 

As in the previous example, the outer orbit of the binary decays with time due to the DF,  such that the rate of stellar encounters steadily increases.  Simultaneously with that,  the inner semi-major axis steadily grows to $\sim 200$ AU, while the periastron distance decays to $\sim 10$ AU. The corresponding eccentricity growth is accomplished predominantly through stellar flybys; even though cluster tides are operating effectively ($\epsilon_\mathrm{GR}\ll 20$), they change $e$ only at the level of $\sim 0.1$ since $\Theta_\mathrm{io}$ does not get low enough, see panel (e). 

At $t\approx 0.08$ Gyr, still far from the centre of the cluster, the binary experiences a strong penetrative encounter with a cluster star ($Q=0.014a$, $v_{\rm rel,\infty}=32$ km s$^{-1}=3.3\sqrt{G\mb/a}$), which completely unbinds it. This happens when $a$ is still substantially below $10^3$ AU, which is our threshold for following its evolution.   Such impulsive ionization is a rather frequent evolutionary outcome given the large initial $a$ that we adopt in these examples (our reasoning behind that is explained above in \S\ref{section:examples}), making the binary 'soft' from the start.

\subsubsection{Example 3: binary experiencing an exchange interaction and then merging}
\label{sec:ex3}

A binary can also experience an exchange interaction whereby, in the course of a close encounter with a cluster star,  one of the binary components gets ejected and the perturber takes its place, see Fig.~\ref{fig:exchange}.

In this example the binary again sinks to the cluster centre due to the DF within $0.2$ Gyr.  In the process, its eccentricity shows a substantial evolution (panel (b)) as it gets increased to $1-e\sim 10^{-4}$ around $t=0.08$ Gyr by encounters with cluster stars (we verified that).  Once the binary has reached a high eccentricity (i.e. low angular momentum), it becomes easy for any given encounter to increase the eccentricity even further,  although this process is stochstic.  After reaching this peak, $e$ then decreases to $\sim 0.1$, again, due to encounters.  Cluster tides play some, but not decisive,  role in this eccentricity evolution.

Upon reaching the cluster center, around 0.18 Gyr, a close encounter with a star ($Q=0.088a$, $v_{\rm rel,\infty}=\SI{4.8}{km/s}=0.11\sqrt{G(m_1+m_2)/a_b}$)  takes place, unbinding one of the black holes and binding another one to the incoming star, thereby forming a new $10 M_\odot + 1 M_\odot$ binary.  This new binary has a rather small semi-major axis $<0.1$ AU (so that the GR precession completely suppresses secular effect of cluster tides), which is necessary to unbind a heavy BH companion, and it eventually merges due to the GW emission at $t\approx 0.7$ Gyr.  

Even though this new binary sits right at the cluster centre where the stellar density is high, the rate of close stellar encounters capable of changing its $a$ drops dramatically.  This is because ${\cal R}\propto a^2$ and the close stellar flybys are very rare, see panel (a).  At the same time, encounters are still capable of temporarily dislodging the new binary from the cluster centre by $\sim 0.2$ pc from time to time,  see panel (f).

\begin{figure}
\includegraphics[width=0.49\textwidth]{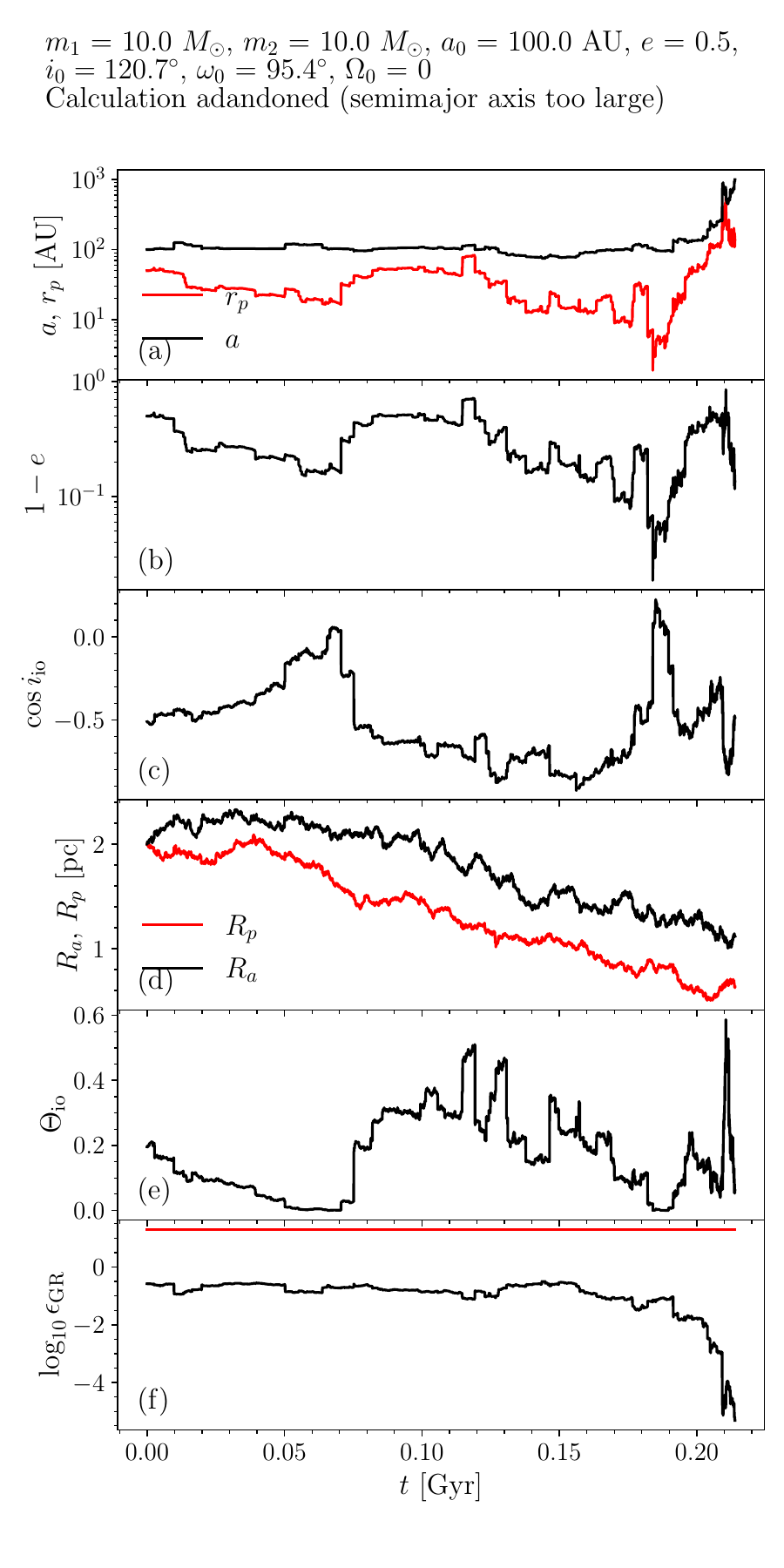}
\vspace{-0.8cm}
\caption{
Same as Fig.~\ref{fig:merged} except the binary inner semimajor axis $a$ eventually exceeds $10^3$ AU, at which point we abandon the calculation and consider the binary to be effectively disrupted. See \S\ref{sec:ex4} for a discussion. }
\label{fig:abandoned}
\end{figure}


\subsubsection{Example 4: binary ionized diffusively}
\label{sec:ex4}

Fig.~\ref{fig:abandoned} shows an example of a binary that is disrupted (ionized) in a diffusive fashion.  As the binary sinks towards the cluster centre due to dynamical friction, it gets softened by encounters that become more frequent as $R$ decreases. Eventually, $a$ crosses $10^3$ AU at which point we stop the calculation since it would get fully unbound soon after anyway.  Here binary ionization occurs as a result of a large number of small changes of $a$ that accumulate in a diffusive fashion, unlike the  case shown in Fig.~\ref{fig:destroyed}, in which a single close encounter disintegrated the binary when its $a$ was still not too far from the hard-soft boundary. 

We also note that in this example, due to the large $a$, cluster tides enable somewhat stronger $e$ evolution than in the previous examples,  with the corresponding $\Delta e\sim 0.2$.

\begin{figure}
\includegraphics[width=0.49\textwidth]{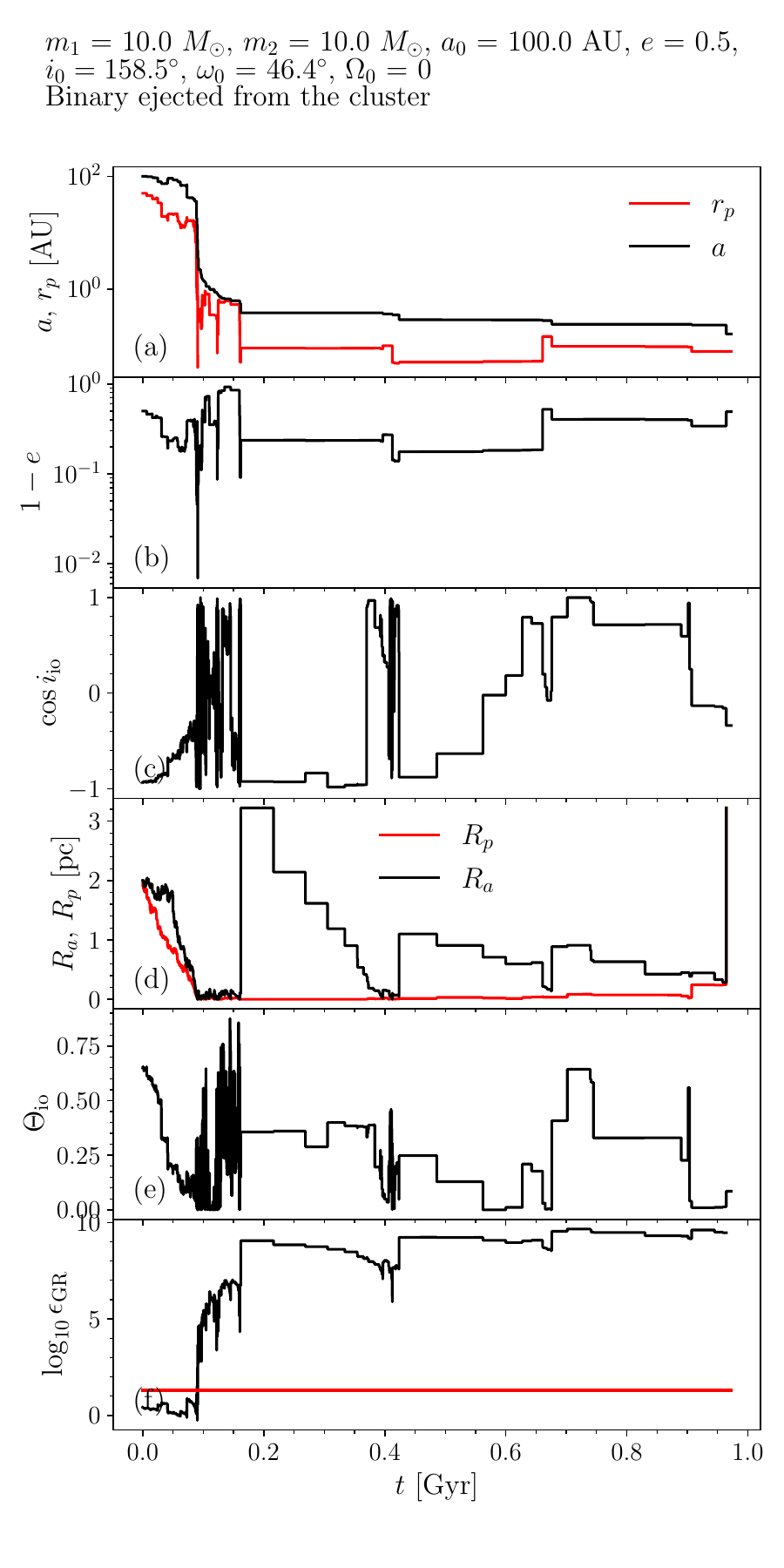}
\vspace{-0.8cm}
\caption{
Same as Fig.~\ref{fig:merged}, except the total cluster mass is $M_\mathrm{cl}=10^5\msun$ and the binary ends up being ejected from the cluster.  See \S\ref{sec:ex5} for a discussion. }
\label{fig:ejected}
\end{figure}


\subsubsection{Example 5: binary ejected}
\label{sec:ex5}

One of the possible consequences of the close encounters is that the binary is sometimes ejected from the cluster. In our simulations that almost always happens in Hernquist potentials with low mass $M_{\rm cl}=10^5\msun$. One of those cases is shown in Fig.~\ref{fig:ejected}.  Here the binary sinks to the cluster centre rather quickly, within 0.1 Gyr, where its $a$ get rapidly reduced by encounters to below 1 AU (effectively turning off cluster tides, see panel (e)). A strong encounter around 0.16 Gyr then kicks the binary into a very radial orbit inside the cluster with $R_p\approx 3$ pc but $R_a\approx 0$ pc (and also increases its $e$). The outer radius $R_p$ then decays by the DF and by 0.4 Gyr the binary orbit is again fully confined to the central regions of the cluster.  This cycle repeats around 0.42 Gyr and 0.7 Gyr, until a particularly strong encounter completely ejects the binary from the cluster around $t=1$ Gyr. 

The exact value of the semimajor axis at which the binary becomes prone to ejections can be estimated as follows. For $m_1\approx m_2$, $m_3\lesssim \mb$ the typical velocity of the cluster star $m_3$ after a close flyby ($Q\sim a$) is $v_{\rm ej,3}\approx\sqrt{G\mb/a}$ \citep[][Eq. 8.21]{degn} and therefore the typical binary recoil velocity is
\eq{
v_{\rm ej,bin}=v_{\rm ej,3}\frac{m_3}{\mb}\approx\sqrt{\frac{Gm_3^2}{\mb a}}.
}
The escape velocity from the centre of a Hernquist potential is $v_{\rm esc} = \sqrt{2GM_{\rm cl}/b}$ \citep[][Eq. 15]{hernquist}. Thus, $v_{\rm ej,bin}>v_{\rm esc}$ and the binary can get ejected from the cluster centre when its $a$ drops below
\eq{
a_\mathrm{ej} \approx b\frac{m_3^2}{2\mb M_{\rm cl}} \approx \SI{0.1}{AU}\frac{b}{\SI{2}{pc}}\qty(\frac{M_{\rm cl}}{10^5\msun})^{-1}\qty(\frac{\mb}{20\msun})^{-1}\qty(\frac{m_3}{\msun})^2.
\label{eq:aej}
}
This estimate is consistent with Fig.~\ref{fig:ejected} since $a$ drops to around 0.1 AU right before the ejection, while $a$ is somewhat larger than $a_\mathrm{ej}$ during the earlier sharp increases of $R_a$ that did not result in an ejection.  Equation (\ref{eq:aej}) also agrees with the results on binary ejections obtained using CMC \citep[][see their Figs. 6]{ejected}. 

The ejection probability is higher in a Hernquist potential than in a cored potential (e.g. Plummer), which is not surprising: in a Hernquist potential $v_{\rm esc}$ from the cluster centre is finite,  while the central stellar density diverges (unlike the Plummer case), so the rate of encounters able to eject the binary increases dramatically (even for small $a$) as the binary sinks into the centre. However, if $a$ is too small, an encounter with $Q\sim a$ capable of ejecting the binary may not happen at all before the binary merges due to the GW emission. This is what we observe in the example shown in Fig. ~\ref{fig:exchange}, in which after the exchange $a$ is significantly below $a_\mathrm{ej}$ (in this example $v_{\rm esc}$ is also higher due to higher $M_{\rm cl}$). Thus, ejections are more likely for $a$ just below $a_\mathrm{ej}$, but not much below.

\begin{figure}
\includegraphics[width=0.49\textwidth]{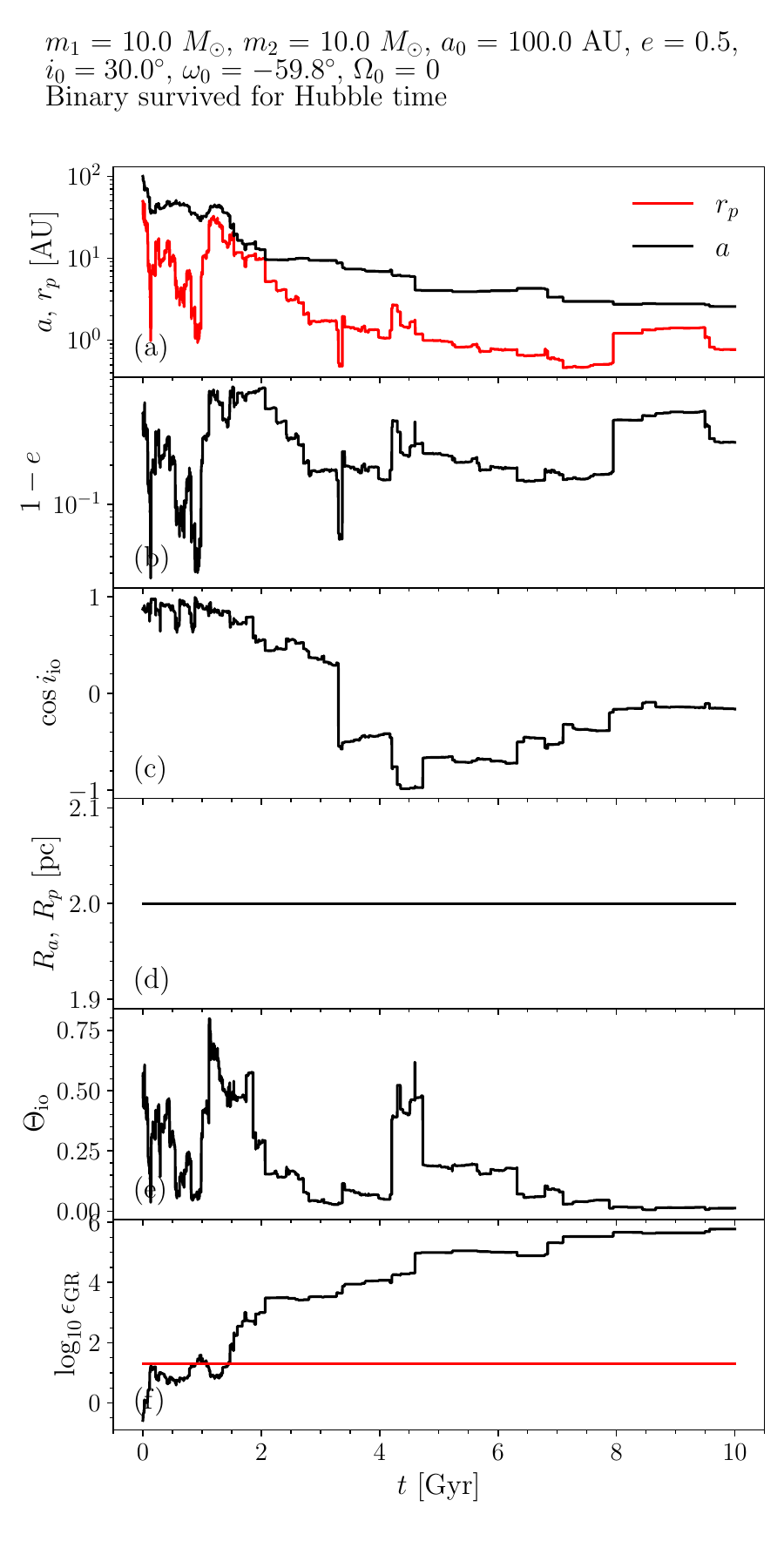}
\vspace{-0.8cm}
\caption{
Same as Fig.~\ref{fig:merged} except the binary baricenter velocity kicks due to encounters are disabled (i.e. no dynamical friction, the (circular) outer orbit of the binary is unchanged, see panel (d)) and the binary survives for 10 Gyr.  See \S\ref{sec:ex6} for a discussion.}
\label{fig:nokicks}
\end{figure}


\subsubsection{Example 6: binary survives for a Hubble time (with kicks disabled)}
\label{sec:ex6}

To demonstrate the impact of DF on the fate of the binary, we performed a set of simulations where we do not take into account the velocity kicks received by the binary CoM as a result of stellar flybys.  This artificially eliminates DF,  but all the inner orbital element changes due to encounters are still fully accounted for.  In such runs the outer orbit stays unchanged at $R=2$ pc resulting in much lower encounter rates than in the runs including DF (see previous examples). 

In Fig.~\ref{fig:nokicks} we show one such run in which a binary gets gradually hardened over time, but its $a$ evolves slowly due to the infrequent flybys, so that the binary survives for 10 Gyr. This almost never happens when the binary CoM recoil is properly accounted for giving rise to DF,  emphasizing the importance of encounter-driven evolution of the outer orbit for determinig the overall evolutionary outcome. 


\subsubsection{Example 7: cluster tides-dominated evolution (with kicks disabled)}
\label{sec:ex7}

In the previous examples, the encounters affected the binary evolution stronger than the cluster tides -- i.e.  the eccentricity oscillations in the absence of encounters described in  \S\ref{section:SAvsDA} are severely washed out by the stochastic changes of $e$ due to encounters.  In part, this is because in none of these examples the binary inner and outer orbital planes were kept orthogonal, which is necessary for large amplitude oscillations of $e$ \citep{paper2}. Also, encounters very easily conceal the signs of smooth cluster tide-driven evolution in time series of $e$.  We will investigate the relative contribution of the two effects in a more quantitative manner in a subsequent work (Rasskazov \& Rafikov 2023, in preparation). 

\begin{figure}
\includegraphics[width=0.49\textwidth]{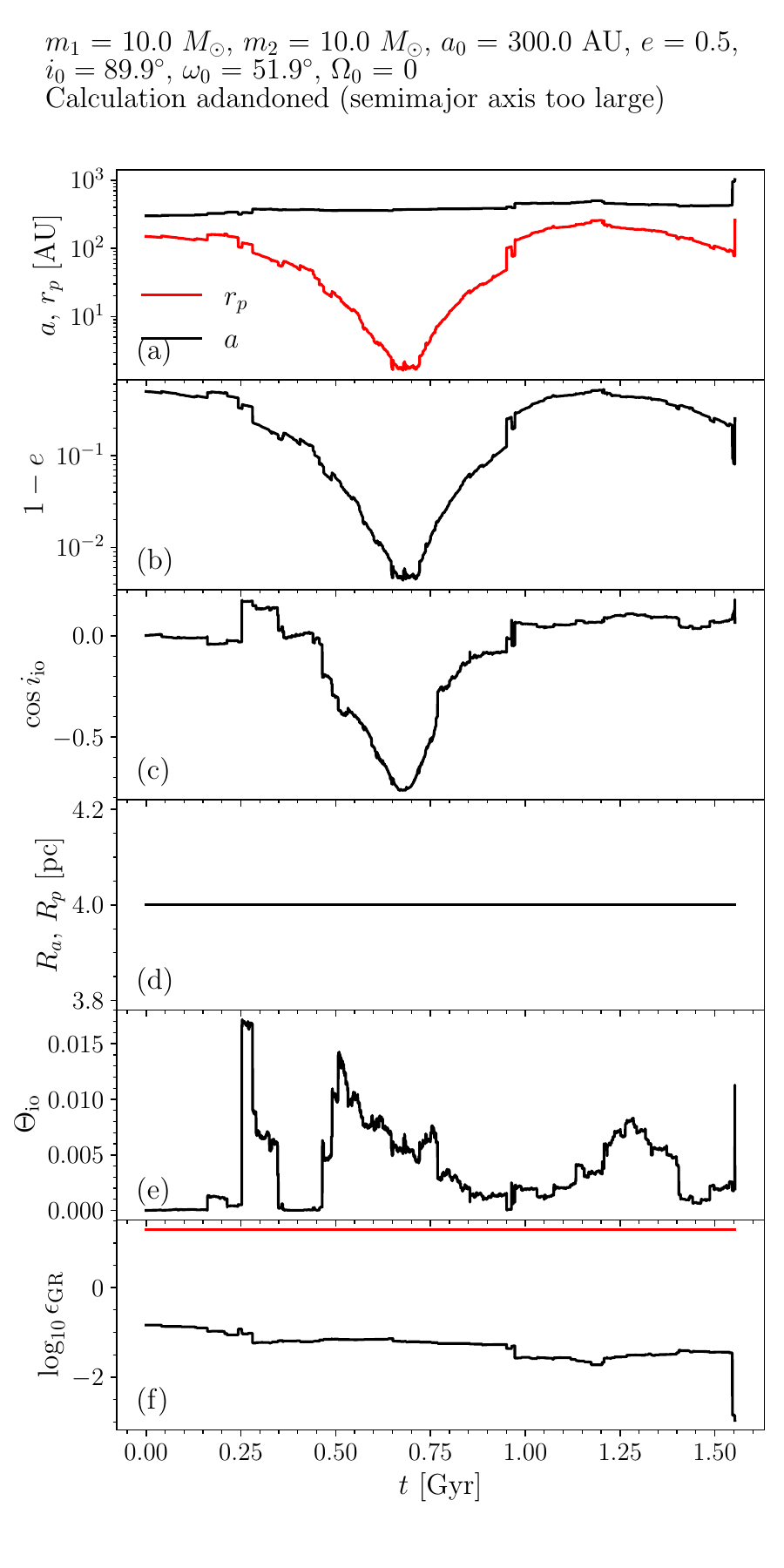}
\vspace{-0.8cm}
\caption{
Same as Fig.~\ref{fig:merged}, but now showing the evolution of an initially wide ($a=\SI{300}{AU}$) and distant ($R_a=R_p=\SI{4}{pc}$) binary in a Hernquist potential with total mass $M_\mathrm{cl}=10^5\msun$ and radius $b=1$ pc with the velocity kicks due to encounters disabled. This example illustrates the possibility of cluster tides dominating the eccentricity evolution (eventually the binary is ionized). See \S\ref{sec:ex7} for a discussion. }
\label{fig:tidalDominated}
\end{figure}

For now,  we just show in Fig.~\ref{fig:tidalDominated} a particular example with the specially designed initial conditions, which allow cluster tides to reveal themselves rather clearly. To achieve that, we put the binary on a rather distant orbit in the cluster ($R_a=R_p=4b$) and artificially maintain this outer orbit by turning off DF (like in \S\ref{sec:ex6}). We have also assumed a higher initial semimajor axis $a=\SI{300}{AU}$ and the initial inclination of $i_0=89.9^\circ$ to enable significant eccentricity oscillations.  This high $a$ prevents GR precession from suppressing the cluster tides, see panel (f).  

One can see that $1-e$ gradually decreases to $\lesssim 10^{-2}$, a behavior that reverses around 0.75 Gyr in an almost symmetric fashion. The $e(t)$ dependence is rather smooth and is only weakly perturbed by encounters, which are rather weak and infrequent so far from the cluster centre (but are still capable of gradually softening the binary). This behavior is strongly indicative of the secular cluster tide-dominated evolution, similar to Lidov-Kozai cycles, which is triggered in this case by the low value of $\Theta_\mathrm{io}\lesssim 10^{-2}$ throughout the evolution. This conjecture is also confirmed by the estimate of the secular timescale expected\footnote{Such a distant binary can be approximated as orbiting in a potential of a point mass $M_\mathrm{cl}$, for which the Lidov-Kozai timescale can be computed using \citet{Kiseleva1998}.} for such a binary, $t_\mathrm{sec}\approx 1.6$ Gyr, which is in good agreement with the approximate oscillation half-period of 0.7 Gyr in Fig.~\ref{fig:tidalDominated}a,b. This binary eventually crosses $a=10^3$ AU threshold at which point we stop our calculation.

\section{Discussion} 
\label{sec:disc}


This study is primarily devoted to the description of a newly developed code BESC and to illustrating its performance via some illustrative examples.  For that reason, here we are not providing quantitative results on binary dynamics except for the preliminary statistics of the outcomes in \S\ref{sec:stat_outcomes} below.  

BESC enables a statistical analysis of the evolution of large samples of binaries in stellar clusters by following individual trajectories of each binary in great detail. The initial parameters of these samples should be provided to BESC as inputs since it does not follow the production of binaries in clusters. As we showed already in \S\ref{section:SAvsDA}, the evolution of an individual binary is typically very chaotic, especially in presence of encounters, see \S\ref{section:examples}.  As a consequence, meaningful conclusions about binary evolution in stellar clusters are possible only in a statistical sense, based on large samples of simulated binaries.  This is the approach we will adopt in our followup work. 

Although the examples that we provided in this study are limited to BH-BH binaries, BESC can certainly be applied to studying other exotic binaries encountered in stellar clusters --- blue stragglers, X-ray binaries, etc.  One just needs to adopt the relevant physical ingredients, as discussed in \S\ref{sec:future} below.


\subsection{Preliminary statistics of the outcomes}
\label{sec:stat_outcomes}


A detailed analysis of the statistics of binary evolution outcomes based on BESC calculations with realistic inputs (stellar mass spectrum, distribution of initial semimajor axes and eccentricities, characteristics of the outer orbits) is left for future work. Nevertheless, we can provide a preliminary (rather low number) statistics of the outcomes at least for the initial conditions used in this work, see \S\ref{section:examples}.  To this goal, we have performed 8 sets of simulations with these initial conditions, about 100 realizations for every combination of the Hernquist and Plummer (not shown earlier) potentials,  cluster mass $M_\mathrm{cl}=10^5\msun,10^6\msun$, with and without the effects of encounters on the outer orbit (i.e. with or without the DF, see \S\S\ref{sec:ex6}, \ref{sec:ex7}). 

Based on this sample, we found that the fraction of binaries that merge within a Hubble time is higher in Hernquist models than in the Plummer ones: 76\% vs 34\%  for $M_\mathrm{cl}=10^5\msun$,  and 45\% vs 19\% for $M_\mathrm{cl}=10^5\msun$. Turning off flyby kicks on the outer orbit (i.e.  preventing the outer orbit from evolving and reaching high stellar density in the cluster center) reduces the merged fraction to around 6\% regardless of other parameters.  This shows that DF is an extremely important factor in determining the fate of the binary: it allows the systems starting far from the cluster centre to eventually merge as the DF brings them into the much denser central regions where they can experience 3-body hardening more efficiently.  None of the binaries end up being disrupted either impulsively (\S\ref{sec:ex1}) or diffusively (via exceeding $a_{\rm in}=10^3$ AU, \S\ref{sec:ex4}) in $M_\mathrm{cl}=10^5\msun$ clusters.  However, in $M_\mathrm{cl}=10^6\msun$ clusters more than half of the binaries get ionized either way.  This is because for our choice of the initial $a=10^2$ AU binaries are hard in the former and soft in the latter cases,  see Fig.~\ref{fig:timescales}b,c.  Finally, as mentioned in \S\ref{sec:ex5},  we observe binaries being ejected from the cluster in significant numbers (18\%) only in a Hernquist model with  $M_\mathrm{cl}=10^5\msun$.


\subsection{Comparison with existing results}
\label{section:comparison}


To the best of our knowledge, BESC is the first code that evolves the binary inner orbital elements taking into account not only the stochastic effects of close stellar encounters but also the secular effect of cluster tides, while also self-consistently evolving the outer orbit of the binary CoM in the cluster (directly capturing the effect of dynamical friction). Earlier studies focused on just one physical process, e.g. \citet{HamersTremaine}, \citet{distantEncountersEffect} and \citet{cmc} include only encounters, while \citet{paper3} and \citet{BubPetrovich} considered only cluster tides. BESC incorporates these effects in a semi-analytical fashion, making it more computationally efficient than the direct N-body calculations \citep[e.g.][]{Li2023}, which naturally account for both effects (as long as they are able to reliably treat close encounters).

Until now, one of the most comprehensive treatments of the binary evolution in stellar clusters has been provided by the Cluster Monte-Carlo (CMC) code developed by \citet{cmc}. We summarize the differences between the CMC and BESC as follows: 
\begin{itemize}
\item Unlike BESC, CMC does not account for cluster tides in the evolution of the inner orbit.
\item CMC ignores the effect of distant flybys ($Q> 2a$) on the inner binary parameters, while BESC explicitly accounts for them in a semi-analytical fashion. 
\item CMC directly includes the effect of very close flybys ($Q< 2a$) on both the inner and outer binary orbits, but for the more distant encounters ($Q> 2a$) it accounts for their effect only on the outer orbit, in a prescribed fashion. 
\item Therefore, CMC should account for the dynamical friction, although \citet{cmc} do not explicitly mention this; BESC accounts for the dynamical friction directly, by explicitly calculating the outcome (change of the CoM motion) of each encounter.
\item CMC follows only the energy and angular momentum of the outer orbit but not its actual shape; the phase of the outer orbit is randomly sampled at each time step. BESC fully traces the outer orbit as a function of time, which is important in aspherical clusters.
\item Since CMC describes every member of the cluster, it accounts for the self-consistent evolution of the cluster as a whole, whereas BESC assumes a prescribed model for the cluster density (which may however be made time-dependent). 
\end{itemize}

These differences of BESC with CMC, as well as with other existing numerical approaches to binary evolution in clusters, are also summarized in Table~\ref{table:comparison}.

 
\subsection{Future extensions}
\label{sec:future}


The implementation of BESC described in this paper makes a number of approximations regarding the setup and relevant physics, many of which can be easily relaxed in the future.

For example, in this paper we have not considered non-spherical potentials, but the cluster tide module implemented by \citet{BubPetrovich} which is used in BESC works also in fully triaxial potentials. We assumed that all perturbers have mass $m_3=1\msun$ but the code can naturally account for a mass spectrum of perturbers (the effect of varying $m_3$ will be explored in Rasskazov \& Rafikov, in prep.). We assume the cluster potential/density distribution to be constant in time, not accounting for cluster relaxation, core collapse, mass segregation, etc. In the future, these effects can be modeled by evolving the cluster mass distribution and potential in a prescribed manner.

Since our focus in this paper is on BH-BH binaries, we accounted for the GR precession and the dissipative effect of the GW emission. However, BESC can be easily extended to treat the phenomena including less relativistic binaries in clusters, for example normal stellar binaries, stars with planets, blue stragglers, X-ray binaries, etc. In such cases one would need to include additional sources of precession --- rotationally- and tidally-induced quadrupoles --- and consider the possibility of tidal dissipation inside the binary components instead of the GW emission. 


\section{Summary}
\label{section:conclusions}


We developed a numerical framework --- BESC --- that allows one, for the first time, to simulate evolution of a binary in a stellar cluster while simultaneously including a variety of physical effects: stellar flybys, tidal forces from the smooth cluster potential, and GR effects. Effect of cluster tides are accounted for at the singly-averaged level, without averaging over the outer orbit of the binary (which we compute explicitly). Moreover, when accounting for stellar flybys we incorporate the effects (on the inner and outer orbit of the binary) of both the close stellar encounters (via the direct 3-body integration of an encounter) as well as the distant flybys (using a semi-analytic secular approximation).

We then carried out a number of binary evolution calculations with various parameters, and our preliminary results can be summarized as follows.
\begin{itemize}
\item In the absence of encounters, we performed a comparison of the singly-averaged (SA) framework for the cluster tide-driven evolution with the existing results \citet{paper4} based on the doubly-averaged (DA) approximation. We found that the binary eccentricity oscillations on the outer orbital timescale can cause binary evolution to deviate from the DA approximation in an unpredictable manner at very high eccentricities $e\to 1$. As a result, in most cases the binary reaches higher eccentricities and merges sooner than in the DA calculations.
\item Encounters with cluster stars play a very important role in a variety of ways. In particular, they tend to disrupt the smooth secular evolution driven by the cluster tides adding a strong chaotic component to the variation of binary orbital elements.
\item Another key outcome of stellar encounters is the dynamical friction that causes the outer orbit of the binary to decay towards the cluster center. This effect is directly accounted for in BESC thanks to its inclusion of the distant encounters.
\item Cluster tides are most effective when the binary semi-major axis is in the range $a\sim 10^2-10^3$ AU, however, these binaries can be easily ionized by passing stars. Also, for lower $a$ the GR precession presents an important obstacle to reaching $e\to 1$ in the course of cluster tide-driven evolution.
\item Cluster tides can dominate binary eccentricity evolution  (over the encounters) only when the binary is very far from the cluster centre and evolution of its outer orbit due to the dynamical friction is suppressed. Nevertheless, as we demonstrate more quantitatively in the future (Rasskazov \& Rafikov 2023, in prep.), the effect of cluster tides on the binary eccentricity evolution cannot be ignored.
\item In agreement with other studies \citep[e.g.][]{ejected}, we sometimes observe exchange interactions or the binaries escaping the cluster.
\item The BH binary merger rate seems to be mainly determined by the mean density at the outer orbit of the binary, which is higher in the Hernquist (cusped) potential compared to the Plummer (cored) potential. Stellar encounters play the key role in promoting mergers in part via the encounter-driven dynamical friction that sinks the binary towards the cluster center, where the stellar density is high.  
\end{itemize}

We also note that BESC can be used to follow the evolution of other binary systems within a cluster, such as hot Jupiters \citep[improving on the numerical method of][]{HamersTremaine}, blue stragglers, X-ray binaries, etc.

\section*{Acknowledgements}

We thank Chris Hamilton, Adrian Hamers and Mathew Bub for helping us with their numerical methods, and also Steven Rieder and Eugene Vaslilev for useful discussions.
{\it Software:} Matplotlib \citep{matplotlib}, numpy \citep{numpy}, astropy \citep{astropy}, galpy \citep{galpy}, ARCHAIN \citep{archain}, binary-evolution \citep{BubPetrovich}, flybys \citep{distantEncounters}. A large part of the long term simulations were performed on the HPC cluster FAWCETT at DAMTP, University of Cambridge. Running our code on FAWCETT would not be possible without the help of the DAMTP research software engineer Irina Davydenkova.

\bibliographystyle{mnras}
\bibliography{bib}

\appendix


\section{Choices of BESC parameters and the accuracy of hybrid integration}
\label{appendix}


\begin{figure}
	\centering
	\subfigure{\includegraphics[width=0.49\textwidth]{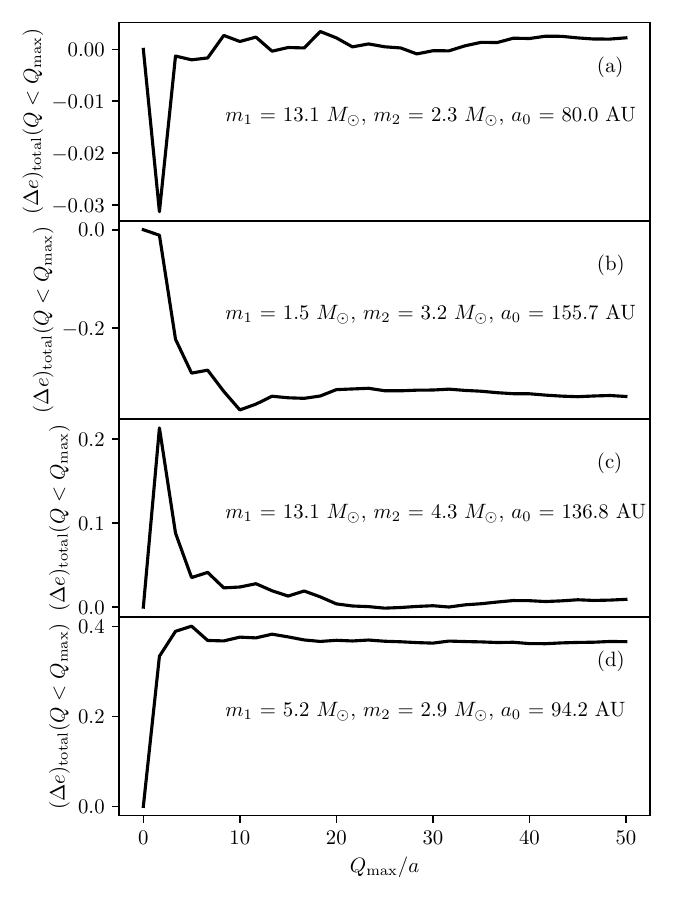}}
 \vspace{-0.7cm}
\caption{
Cumulative change of the binary eccentricity $(\Delta e)_\mathrm{total}$ that it has experienced in its lifetime due to stellar encounters with periastron distances $Q<Q_{\rm max}$, plotted as a function of $Q_{\rm max}/a$ for several representative binary evolutionary tracks. In these tracks the effect of encounters on the outer orbit has been turned off (i.e. no dynamical friction), the outer orbit lies between $R_p=1.9$ pc and $R_a=2.1$ pc. Calculation assumes Hernquist potential with total mass $M_\mathrm{cl}=10^6\msun$ and scale radius $b=2$ pc. The initial binary inclination and eccentricity are $i=89.8^\circ$ and $e=0.5$, respectively. One can see that $(\Delta e)_\mathrm{total}$ saturates for $Q_{\rm max}/a\gtrsim 20$, motivating our choice of $Q_{\rm max}=50a$ in BESC, see \S\ref{section:changes_dist}.
} 
\label{fig:de_total(Q_max)}
\end{figure}

We start by providing a more quantitative justification for our choice of the upper limit $Q_\mathrm{max}=50a$ for an encounter to be considered at all. Figure \ref{fig:de_total(Q_max)} shows, for several typical binary evolutionary tracks (in different panels), the total change in eccentricity $(\Delta e)_\mathrm{total}$ over the entire binary lifetime due to the flybys with pericentre distances $Q<Q_{\rm max}$ only (i.e., with more distant flybys excluded), as a function of $Q_{\rm max}/a$. We see that $(\Delta e)_\mathrm{total}$ typically saturates and becomes essentially independent of $Q_{\rm max}$ at $Q_{\rm max}/a\sim 20$. In other words, accounting for the encounters more distant than $Q_\mathrm{max}=50a$ would have almost no effect on binary eccentricity evolution, with good margin.

\begin{figure*}
    \includegraphics[width=\textwidth]{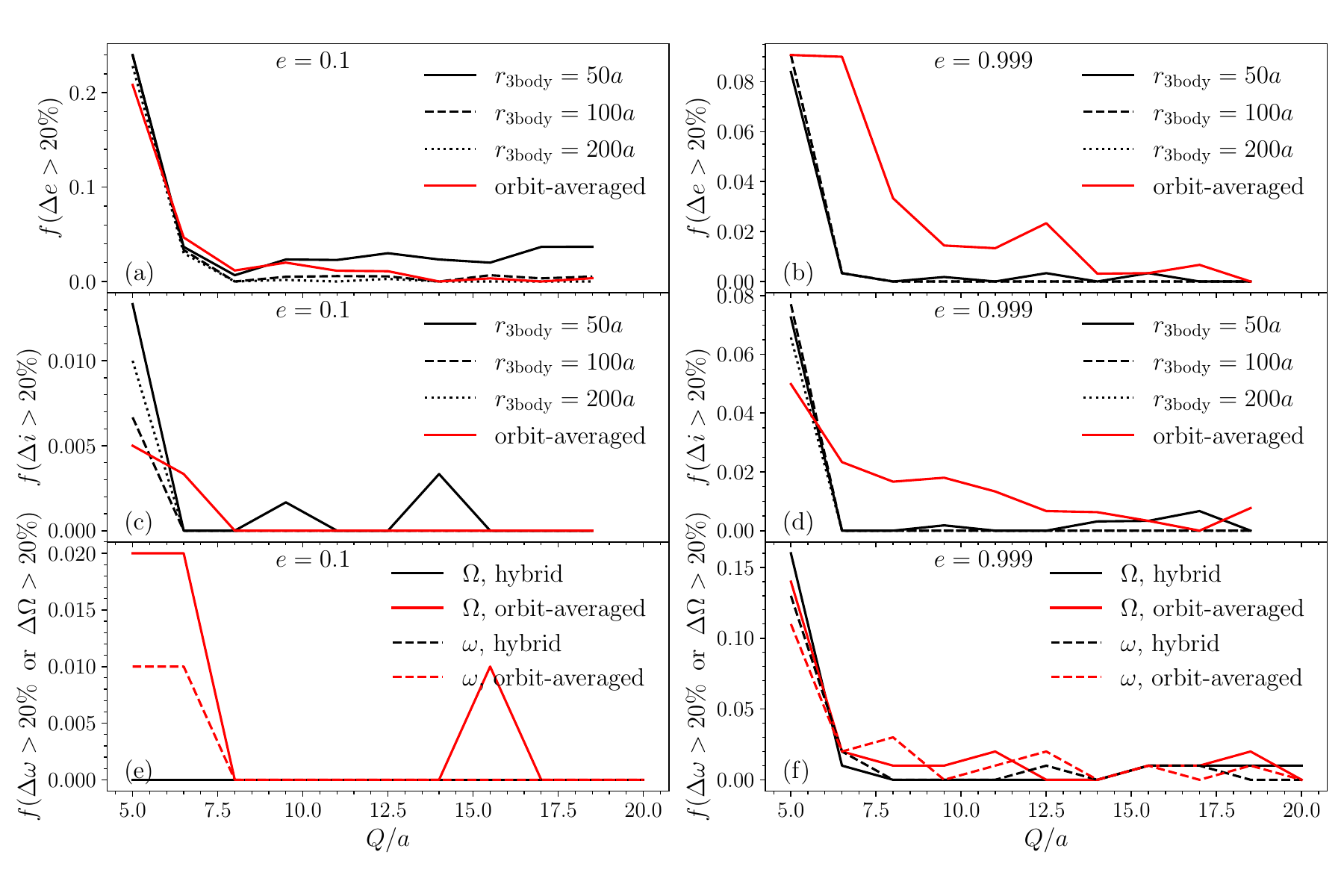}
 \vspace{-1cm}
\caption{
Fraction $f$ of stellar flybys (for initial conditions randomly chosen as described in Appendix \ref{appendix}) for which the hybrid (black) and orbit-averaged (red) calculations of the orbital elements changes in an encounter --- (top) eccentricity, (middle) inclination, (bottom) longitude of the ascending node and the argument of periapsis  --- result in $>20\%$ deviations relative to the exact 3-body integration. Initial binary eccentricity is $e=0.1$ (left) and $e=0.999$ (right). Hybrid calculations in panels (a)-(d) use different values of $r_\mathrm{3body}$ indicated in panels; the bottom row assumes $r_\mathrm{3body}=100a$.
} 
\label{fig:outliers}
\end{figure*}

\begin{figure}
    \includegraphics[width=0.5\textwidth]{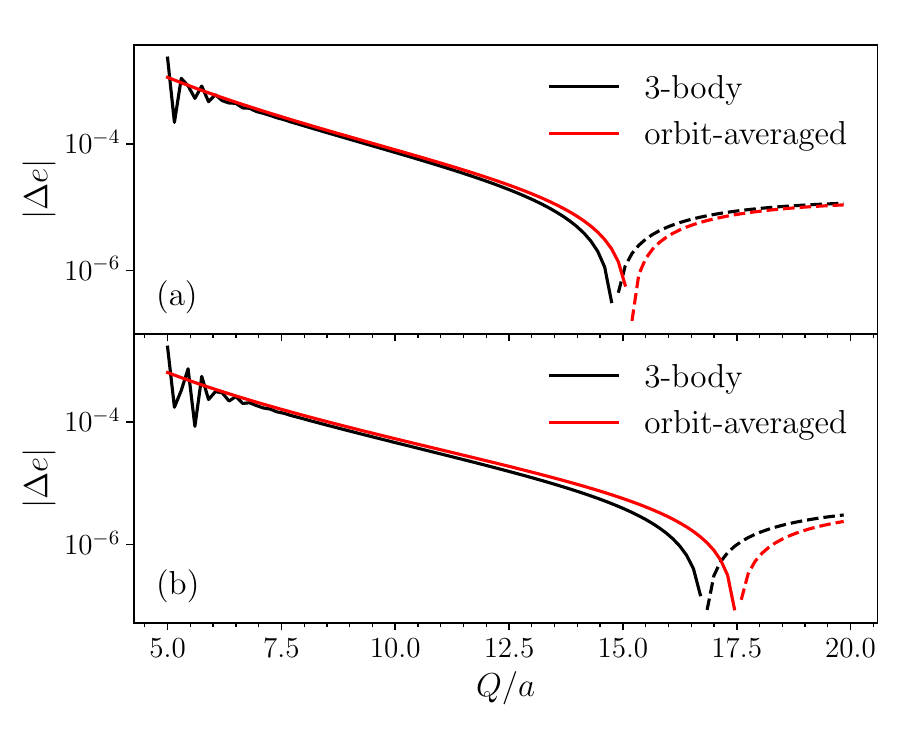}
 \vspace{-0.8cm}
\caption{
Two examples of the dependence of the eccentricity change $\Delta e$ on the perturber pericentre distance $Q$ calculated using the 3-body (black) and the orbit-averaged (red) methods, all other parameters being equal, with $\Delta e$ changing sign at some $Q$. The solid (dashed) lines are for $\Delta e>0$ ($<0$). In both examples $m_1=m_2=m_3=5\msun$, $a=1$ AU, $v_\mathrm{rel,\infty}=3$ km/s, $e=0.1$, $i=\Omega=\omega=0^\circ$. The perturber parameters are (a) $i=64.7^\circ$, $\Omega=269.3^\circ$, $\omega=232.6^\circ$ and (b) $i=55.0^\circ$, $\Omega=89.4^\circ$, $\omega=246.4^\circ$ 
} 
\label{fig:qdependence}
\end{figure}

\begin{figure}
	\centering
	\subfigure{\includegraphics[width=0.49\textwidth]{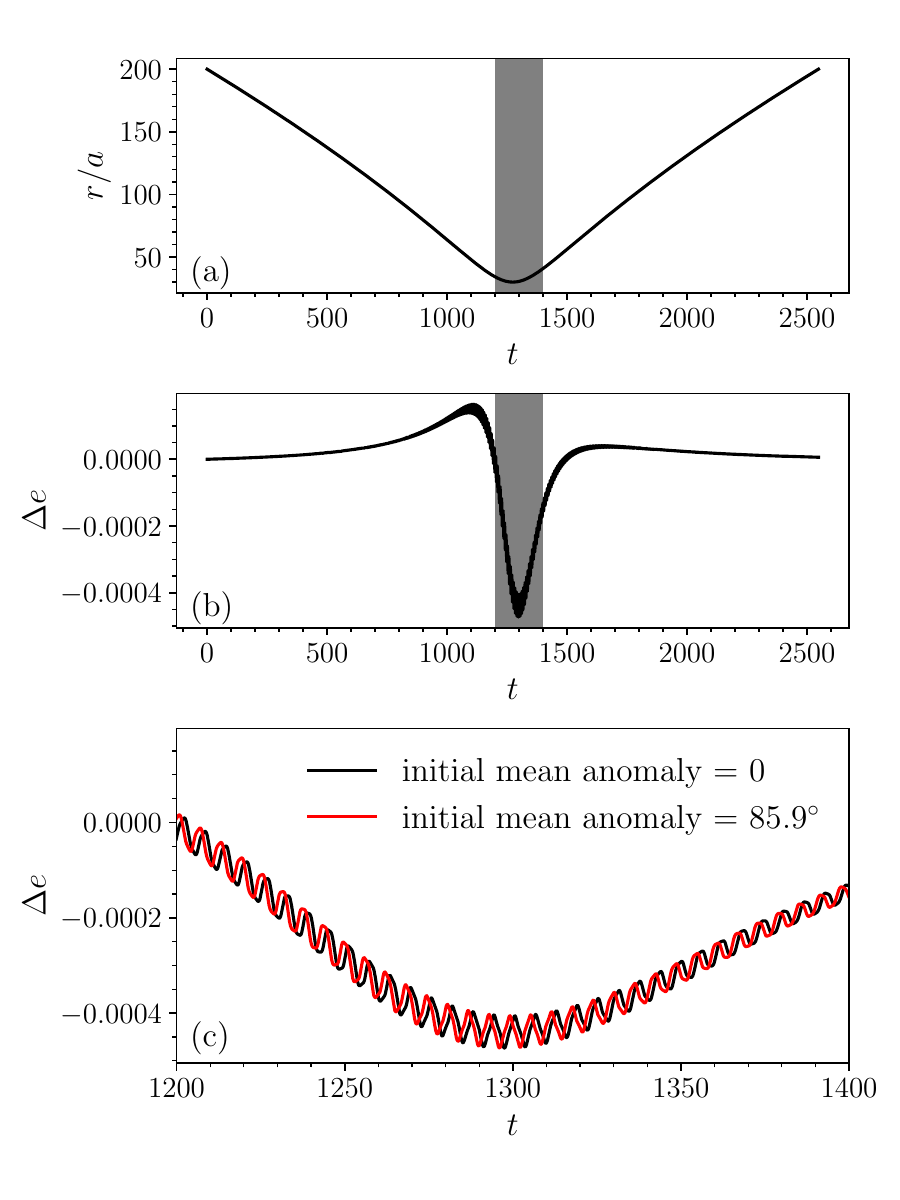}}
 \vspace{-0.8cm}
\caption{
(a) Evolution of the perturber-binary distance $r$ in units of $a$ and (b) binary eccentricity change $\Delta e$ during an encounter, calculated using the direct 3-body integration ($t$ is time in units such that the binary orbital period is $2\pi$). Grey bands show an interval, which is zoomed over in panel (c). Black and red curves use the same parameters except for the different binary initial mean anomaly (0 and $85.9^\circ$ correspondingly). The red line is not shown in the top two panels for the sake of clarity as it almost coincides with the black one.  
}
\label{fig:de(t)}
\end{figure}

\begin{figure}
    \includegraphics[width=0.49\textwidth]{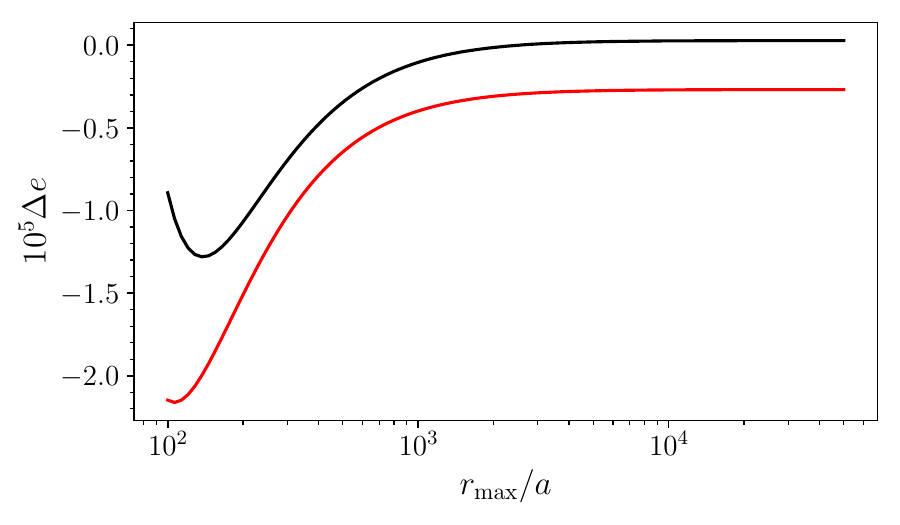}
 \vspace{-0.8cm}
\caption{
Two typical examples of the dependence of the eccentricity change $\Delta e$ during an encounter on $r_\mathrm{max}$ -- the distance between the perturber and the binary where we start/stop the integration. 
} 
\label{fig:rmax}
\end{figure}

Next we describe some tests of accuracy of the orbit-averaged/hybrid approach (described in \S\ref{section:changes_dist}) for calculating the outcome of a stellar flyby and motivate our choices of the parameters $Q_\mathrm{hyb}$, $r_\mathrm{3body}$ and $r_\mathrm{max}$ employed by BESC. We do this by comparing the result of a hybrid calculation with a direct 3-body integration of an encounter and analyzing the resultant deviations. For every value of the encounter pericenter distance $Q$ we sample 100 initial conditions, with all the orbital angles randomized and the rest of parameters set as follows:
\begin{itemize}
\item The masses of all three bodies are $m_1=m_2=m_3=5\msun$
\item Binary semimajor axis is $a=\SI{1}{AU}$ 
\item Binary eccentricity $e=0.1$ or $e=0.999$
\item Initial velocity of the third body (at infinity) $v=\SI{3}{km/s}$
\end{itemize}

For every such initial condition we simulate an encounter and measure the change of the binary orbital elements ($\Delta e$, $\Delta i$ etc.) using all 3 approaches: 3-body, orbit-averaged and hybrid. In Fig.~\ref{fig:outliers} we show the fraction of the simulated encounters where the relative error in $\Delta e$, $\Delta i$ etc. for the orbit-averaged/hybrid approach (assuming 3-body approach is precise) exceeds $20\%$ (e.g. when $\qty|\Delta e_{\rm hybrid}/\Delta e_{\rm 3body}-1| > 0.2$). 

As expected, the orbit-averaged approximation breaks down at $Q/a\lesssim 5$ where the perturbations to both the binary and the perturber's orbit become too strong. For that reason we set $Q_\mathrm{hyb}=10a$, so that all encounters with $Q<Q_\mathrm{hyb}$ are processed using the hybrid approach, including the direct 3-body integration. 

In some cases, there is also a small fraction of significant errors at large $Q/a$. These deviations at large $Q/a$ are due to the $\Delta e(Q/a)$ function going through zero at a certain $Q/a$, so that even very small differences in $\Delta e$ (between the 3-body integration and other methods) lead to large relative errors. This phenomenon is illustrated in  Fig.~\ref{fig:qdependence}, where we plot $|\Delta e|$ for two sets of initial conditions (described in the caption), both for relatively low $e=0.1$; in both cases $\Delta e$ passes through zero at $Q/a\gtrsim 15$.

Even with the hybrid approach, the fraction of discrepancies with the 3-body simulations starts to rise when $Q/a\lesssim6$. However, that happens not because of hybrid approach being inaccurate compared to 3-body, but rather due to the chaotic nature of close interactions, i.e. the strong dependence of the flyby outcome on the initial binary mean anomaly (which is picked randomly in both hybrid and 3-body methods). For the same reason, 3-body integrations with different initial mean anomalies would also disagree with each other. 

At higher $Q/a$ the number of discrepancies goes down, but only when $r_\mathrm{3body}$ --- the distance where we switch from the orbit-averaged to 3-body integration --- is large enough, see Fig.~\ref{fig:outliers}. For high enough $r_\mathrm{3body}\gtrsim 100a$, the fraction of discrepancies falls to zero at $Q\gtrsim 8a$ while it sometimes stays at a nonzero level for lower $r_\mathrm{3body}$. Fig.~\ref{fig:de(t)} illustrates why that happens. It shows the  evolution of (a) perturber separation $r$ and (b) binary eccentricity during an encounter (computed with the 3-body integrator). Zooming in on the moment of the closest approach in panel (c), one can see the eccentricity oscillations on the orbital period of the binary with the amplitude comparable with (or even larger than) the resultant total eccentricity change. In the 3-body approach those oscillations are naturally accounted for, while in the orbit-averaged approach they are averaged out. In the hybrid approach, however, neither of those two things happen, and the initial binary phase (randomly chosen when the perturber separation crosses $r_\mathrm{3,max}$) appears to affect the result more than it actually does. While the $\Delta e$ values obtained using the hybrid approach are still precise on average, their dispersion is artificially increased. This effects weakens as we choose larger $r_\mathrm{3body}$. 

Based on these considerations, in our calculations we set $r_\mathrm{3body}=100a$ and use the orbit-averaged approximation whenever $Q>Q_\mathrm{hyb}=10$, switching to the hybrid one for $Q<Q_\mathrm{hyb}$, as mentioned in \S\ref{section:changes_dist}. That way, the fraction of simulations with $>20\%$ errors in the eccentricity calculation stays below $2\%$.

We start/stop the encounter processing when the distance from the third body to the binary is $r=r_\mathrm{max}=10^4a$. While this appears as a very distant threshold, in some cases we really need to start the hybrid or purely orbit-averaged integration that early to fully account for the effect of a flyby. This is illustrated in Fig.~\ref{fig:rmax} where we show the dependence of $\Delta e$ on $r_\mathrm{max}$ for two sets of initial conditions (with pericentre distances $Q\sim 10a$ in both cases). One can see that in these cases $\Delta e$ converges on its true value only for $r_\mathrm{max}\gtrsim 10^3a$. To be fair, in both these cases $\Delta e$ is quite small (these parameter sets are close to the $\Delta e=0$ points discussed above and illustrated in Fig. \ref{fig:qdependence}) and for most other choices of the encounter initial conditions such high $r_\mathrm{max}$ is not required for an accurate $\Delta e$ calculation. Nevertheless, since increasing $r_\mathrm{max}$ above $10^4a$ does not increase the orbit-averaged/hybrid calculation time significantly, we set $r_\mathrm{max}=10^4a$ in BESC.

\bsp	
\label{lastpage}
\end{document}